\documentclass{sigchi}

\usepackage{balance}       
\usepackage{graphics}      
\usepackage[T1]{fontenc}   
\usepackage{txfonts}
\usepackage{mathptmx}
\usepackage[pdflang={en-US},pdftex]{hyperref}

\usepackage{color}
\usepackage{booktabs}
\usepackage{textcomp}
\usepackage{siunitx}
\usepackage{enumitem}

\usepackage{accsupp}
\usepackage{pdfcomment}

\usepackage{microtype}        
\usepackage{ccicons}          
\usepackage[utf8]{inputenc} 

\usepackage{cuted}
\usepackage{capt-of}

\usepackage{todonotes}

\DeclareGraphicsExtensions{%
    .jpg,.png,.PNG,%
    .pdf,.PDF,%
    .mps,.jpeg,.jbig2,.jb2,.JPG,.JPEG,.JBIG2,.JB2}

\def\plaintitle{
CoVR: A Large-Scale Force-Feedback Robotic Interface for Non-Deterministic Scenarios in VR}

\def\emptyauthor{}
\def\plainkeywords{Virtual Reality; Haptics; Kinesthetic Feedback; Actuated device; Tangible User Interface; Encountered-type Haptic Devices; Robotic Graphics}

\makeatletter
\def\url@leostyle{%
  \@ifundefined{selectfont}{
    \def\UrlFont{\sf}
  }{
    \def\UrlFont{\small\bf\ttfamily}
  }}
\makeatother
\def\pprw{8.5in}
\def\pprh{11in}

\definecolor{linkColor}{RGB}{6,125,233}
\hypersetup{%
  pdftitle={\plaintitle},
  pdfauthor={\emptyauthor},
  pdfkeywords={\plainkeywords},
  pdfdisplaydoctitle=true, 
  bookmarksnumbered,
  pdfstartview={FitH},
  colorlinks,
  citecolor=black,
  filecolor=black,
  linkcolor=black,
  urlcolor=linkColor,
  breaklinks=true,
  hypertexnames=false
}

\newcommand\authors[2]{\\[-0.9em]\newcommand\AND[0]{\hspace{#1}}#2}
\newcommand\affiliation[1]{\alignauthor{#1}}

\newcommand\numberofaffiliations[1]{\numberofauthors{#1}}

\def\SYSTEM/{CoVR}

\usepackage{xcolor}

\begin{document}

\title{\plaintitle}
\numberofaffiliations{1}
\author{
\authors{4em}{Elodie Bouzbib$^{1,2}$\AND 
              Gilles Bailly$^{1}$ \AND 
              Sinan Haliyo$^{1}$ \AND 
              Pascal Frey$^{2}$}
  \affiliation{
    \affaddr{$^{1}$Sorbonne Universit\'e, CNRS, ISIR. Paris, France}\\
    \affaddr{$^2$Sorbonne Universit\'e, ISCD. Paris, France}%
    }
}


\maketitle

\begin{strip}\centering 
\pdftooltip{\includegraphics[width=\textwidth]{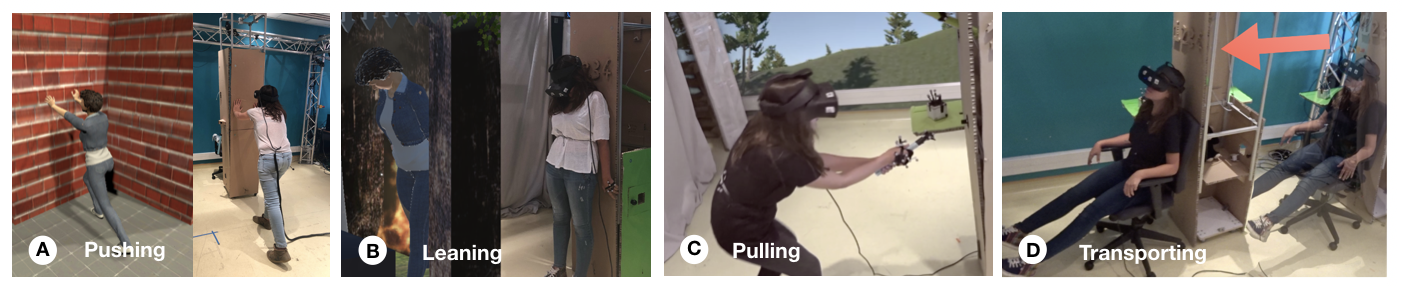}}{This teaser demonstrates the use of CoVR in four different scenarios. In all these scenarios, the user is wearing the Oculus Rift S Head-Mounted Display. On the first one, the user is pushing on one of CoVR's panels, while its virtual avatar pushes on a brick wall. The second picture shows the user and her avatar leaning. The physical version is leaning over CoVR's panels while the avatar is leaning on a chimney (reference to Harry Potter). The third picture shows the user being pulled by CoVR. The background of the picture shows a park/a forest, and reflects the user's viewpoint in virtual reality. It suggests that the user is flying above this forest. The fourth picture shows the user being transported from one location of the room to another one. An arrow shows the displacement of CoVR.}
  \captionof{figure}{\SYSTEM/ is a physical column mounted on a 2D Cartesian ceiling robot to provide strong kinesthetic feedback (> 100N) in a room-scale VR arena. The column panels are interchangeable and its movements can safely reach any location in the VR arena thanks to XY displacements and trajectory generations avoiding collisions with the user. When \SYSTEM/ is static, it can resist to body-scaled users' actions, such as (A) users pushing on a static tangible rigid wall with a high force or (B) leaning on it; When \SYSTEM/ is dynamic, it can act on users. (C) \SYSTEM/ can pull the users to provide large force-feedback or even (D) transport the users.
}
  \label{fig:top_photo}
\end{strip}

\begin{abstract}
We present \SYSTEM/, a novel robotic interface providing strong kinesthetic feedback (100 \si{\newton}) in a room-scale VR arena. It consists of a physical column mounted on a 2D Cartesian ceiling robot (XY displacements) with the capacity of (1) resisting to body-scaled users’ actions such as pushing or leaning; (2) acting on the users by pulling or transporting them as well as (3) carrying multiple potentially heavy objects (up to 80kg) that users can freely manipulate or make interact with each other. We describe its implementation and define a trajectory generation algorithm based on a novel user intention model to support non-deterministic scenarios, where the users are free to interact with \textit{any} virtual object of interest with no regards to the scenarios' progress. A technical evaluation and a user study demonstrate the feasibility and usability of \SYSTEM/, as well as the relevance of whole-body interactions involving strong forces, such as being pulled through or transported.
\end{abstract}

\section{Introduction}

While visual and auditory displays in Virtual Reality (VR) have reached a level where the produced stimuli are quite convincing, haptic technology is still poor compared to the rich ways humans can interact with their environment. Multiple directions have been envisioned to enhance the users' haptic experiences in VR, through hand-held controllers or wearables simulating the environment \cite{de_tinguy_weatavix_2020, benko_normaltouch_2016}, or through the direct manipulation of passive props \cite{insko_passive_2001, bryson_direct_2005}.

In these regards, McNeely introduced Robotic Graphics and more specifically \textit{Robotic Shape Displays} (RSD) \cite{mcneely_robotic_1993} in 1993 as a \textit{concept} for providing force-feedback in VR. It explores the use of a robotic interface in the VR arena to provide haptic experiences, while the user remains unencumbered (no wearables, no controllers) \cite{wexelblat_virtual_1993}. It aims to stay out of reach when no interaction is required, and to displace passive props to dynamically overlay virtual ones otherwise. 

Multiple expanding fields such as Gaming, Training or Simulation could benefit from this concept. Several approaches have already been conducted to instantiate it, from human actuators \cite{cheng_turkdeck:_2015} to drones \cite{abtahi_beyond_2019}. However, they often suffer from several technical trade-offs (e.g. cost, real workspace, embedded mass, speed, accuracy) that grow into interaction challenges (e.g. locomotion whole body interaction, force-feedback, free manipulation of multiple props).


In this paper, we propose \SYSTEM/\footnote{CoVR, pronounced "Cover", stands for a Column in VR which physically covers for virtual objects}, a novel Robotic Shape Display providing whole-body interactions and strong force feedback in room-scale arenas. 
We detail our approach from three complementary perspectives:




From a \textit{mechanical} perspective, \SYSTEM/ is
a 2D Cartesian ceiling robot, carrying a column. With only two single degrees of freedom (XY displacements), \SYSTEM/ is a large-scale grounded robot which exhibits notable mechanical capabilities: high lateral and vertical force feedback and perceived stiffness (over 100~\si{\newton}) and load capabilities (over 80~\si{\kilogram}). It hence can transport a variety of potentially heavy props and objects to stand in for their virtual counterparts. 
We demonstrate that our implementation is fast (over 1.0~\si[per-mode=symbol]{\metre\per\second}) and accurate (under 2~\si{\cm} in a 30~\si{\cubic\meter} arena).


From a \textit{software perspective}, while it is easy to control Cartesian robots as they only rely on XY displacements, defining the robot target positions can be challenging. The system has to predict where the next interaction will occur, while avoiding the user for unexpected collisions. We thus elaborate a user intention model to make CoVR always available prior to the users’ interactions, even in non-deterministic scenarios, i.e. where users are free to interact with \textit{any} virtual object of interest with no regards to the scenarios’ progress.
Given the position and orientation of the user as well as the objects of interest's, the model estimates the best positions for the robot and its optimal trajectories to reach an object of interest while respecting safety constraints (e.g. safe-zones around the user).

A technical evaluation demonstrates that (1) user intentions can be captured using available data from a single HMD with no additional apparatus in a room-scale VR arena and that (2) our system successfully reaches targets prior to interactions in most of non-deterministic scenarios with randomly distributed targets and distractors. 





From the \textit{interaction} perspective, \SYSTEM/ offers (1) body-scaled interactions such as leaning or pushing involving strong forces with a static column (Figure \ref{fig:top_photo} - A, B), or (2) dynamic interactions such as being pulled through by large traction forces (Figure \ref{fig:top_photo} - C). \SYSTEM/ can displace over 80kg of embedded mass, which enables (3) interactions with potentially heavy objects that users can freely manipulate but also (4) transport of the users themselves (Figure \ref{fig:top_photo} - D). 

We report on a user study demonstrating (1) the robustness of both \SYSTEM/ mechanical and software implementations, (2) the benefits of a robotic interface providing body-scale interactions and strong forces and (3) in particular "being transported", which was the favourite one.

\section{Related Work}

Approaches to provide physical interactions in virtual reality are to either simulate physical objects or to exploit the ones available from the environment. \SYSTEM/ is part of a hybrid approach, which uses robotic devices to displace or render physical objects.


\subsection{Simulating physical objects}

Several devices have been proposed to simulate physical objects in VR, with a trade-off between the quality of the haptic rendering, especially kinesthetic feedback and the workspace size \cite{seifi_haptipedia:_2019}. These devices are usually limited to desktop usage \cite{haption_virtuose_2019, force_dimension_force_2019} but can render high quality kinesthetic feedback (over 30N) \cite{seifi_haptipedia:_2019}.
Most of these solutions provide stimulation at the scale of the hand without any regards concerning whole-body interactions.
To widen the workspace, alternatives attach the device to mobile platforms \cite{lee_system_2009,formaglio_performance_2005, satler_control_2011, nitzsche_design_2003, gosselin_widening_2007,lee_mobile_2007, haption_scale1_2019, gosselin_widening_2007}, which remain slow and require to be held continuously whilst moving (which is opposed to Krueger's postulate to develop unencumbered artificial realities \cite{wexelblat_virtual_1993}).

In contrast, several low-cost haptic devices have been proposed in HCI. Typically, wearable or hand-held devices \cite{noauthor_teslasuit_2019, noauthor_cybergrasp_2019,lopes_impacto:_2015,auda_around_2019, benko_normaltouch_2016, whitmire_haptic_2018, choi_claw:_2018, teng_pupop:_2018, heo_thors_2018,strasnick_haptic_2018, lee_torc:_2019, de_tinguy_weatavix_2020, amirpour_design_2019, choi_grabity_2017} 
are naturally compatible with large environments.
They can simulate various haptic features (weight, stiffness, shape, texture) on different body parts. However, they provide limited kinesthetic feedback and they need to be held continuously.


\subsection{Exploiting physical objects} 

The second approach exploits directly physical objects placed in the VR arena, following Insko’s postulate that passive props enhance virtual environments \cite{insko_passive_2001}. For instance, one solution is for each virtual object to \textit{annex} a physical object with similar properties in the VR arena \cite{hettiarachchi_annexing_2016}. Another solution is to use \textit{human actuators} \cite{cheng_turkdeck:_2015} who execute a subtle choregraphy to move the physical objects at the right place and time. The choregraphy is rather costly and time-consuming. Some solutions aim to reduce the number of human actuators to move physical objects by using other users \cite{cheng_mutual_2017, cheng_haptic_2014} or even the users themselves \cite{cheng_iturk:_2018} at the cost of reducing the number of interactive features.

\subsection{Robotic Shape Displays}

The concept of Robotic Shape Displays (RSD) \cite{mcneely_robotic_1993} focuses on the \textit{mobile, unencumbered and untethered aspect of the human} \cite{wexelblat_virtual_1993}. It consists of using a robot to overlay virtual objects with physical props. We distinguish the robotic system (hardware) from their trajectory generation algorithms (software): 

\textit{Robotic system}. Recently, several classes of prototypes aimed to instantiate the original Robotic Shape Display concept by displacing objects \cite{suzuki_roomshift_2020, abtahi_beyond_2019,he_physhare:_2017,kim_encountered-type_2018, hsin-yu_huang_haptic-go-round_2020} or simulating them \cite{siu_shapeshift:_2018} to meet the users without impairing their movements.

Drones can transport objects that users can explore \cite{yamaguchi_non-grounded_2016, knierim_tactile_2017, hoppe_vrhapticdrones:_2018}  or manipulate \cite{abtahi_beyond_2019} in a theoretically unlimited workspace (in practice, the workspace is limited to 2.5 $m^3$ because of technical constraints). This approach only provides a \textbf{small amount of force feedback} as state-of-the-art drones can not resist to human actions. Moreover, this technology has several technical limitations including speed ($<0.5ms/s$), accuracy ($>7cm$), autonomy and safety. Moreover, the accuracy and reliability of these systems decrease with the embedded mass of the props, which reduces the range of available scenarios with high kinesthetic feedback.

Lightweight mobile robots \cite{he_physhare:_2017,suzuki_roomshift_2020, wang_movevr_2020}) or lightweight robotic arms \cite{araujo_snake_2016, vonach_vrrobot:_2017, yokokohji_path_2001, mcneely_robotic_1993, yafune_haptically_2011, shigeta_motion_2007, hoshino_contruction_1995} provide a \textbf{medium amount of force feedback}, but lack an efficiency of displacements. Mobile robots are limited in speed (< 0.5 m/s) and autonomy while robotic arms are limited to desktop usage. 


Only few prototypes provide a \textbf{high amount of force feedback}. TilePop \cite{teng_tilepop:_2019} or LiftTiles\cite{suzuki_lifttiles_2020} modify the floor topology with a large inflatable mat covering the surface of the VR arena. This provides high vertical force feedback and hence supports whole-body interactions below 1m (eg user sitting \cite{zielasko_either_2020}), despite slow inflation (5s) and deflation (20s) times.
Relying on a robust robotic arm such as a Kuka \cite{kim_encountered-type_2018, mercado_design_2020}, can produce a high blocking force and hence provide large kinesthetic feedback at a body-scale, in spite of its cost. This robustness obviously goes along with strong software safety considerations around the user. These prototypes do not easily scale to VR arenas and do not let users freely manipulate a wide range of props.



\textit{Trajectory Generation}. The quality of the interaction does not only depend on the hardware implementation (eg speed) but also on the algorithm to generate trajectories, especially in non-deterministic scenarios\footnote{Non-deterministic scenarios consist of scenarios where multiple virtual objects are available at the same time, and the user is free to interact with any of them. The system does not know in advance which one to physically overlay.}. When the device does not know in advance which virtual object to physically overlay, it is thus necessary to (1) build a user \textbf{intention model} (e.g. \cite{binsted_eyehand_2001}) to estimate what will be next object to overlay and (2) a \textbf{path planning} algorithm (e.g. \cite{cheng_sparse_2017}) to displace the robot to the target location without colliding obstacles. Only few of the above systems \cite{kim_encountered-type_2018, yokokohji_wysiwyf_1999, yokokohji_path_2001} rely on these components to support non-deterministic scenarios. However, they require multiple robots \cite{suzuki_roomshift_2020, vonach_vrrobot:_2017} or specific devices (e.g. eye-tracker \cite{binsted_eyehand_2001}). In contrast, we elaborate a low-computational intention model working with common HMDs.


In summary, Robotic Shape Display systems still face multiple interaction (e.g. whole body interaction, locomotion, force feedback, free manipulation of multiple props) and technical challenges (speed, accuracy, safety, price). 
\SYSTEM/ addresses many of these challenges and focuses on providing large force-feedback (100N).  \SYSTEM/ supports whole-body interactions while letting users physically unencumbered. It is designed for Gaming or Training purposes in large virtual arenas ($\approx$30$m^3$). \SYSTEM/ can carry potentially heavy physical props to match virtual objects the user is about to interact with, without sacrificing speed, accuracy, safety or price.

\section{Design and Implementation of \SYSTEM/}

\begin{figure}[b]
  \centering
  \pdftooltip{\includegraphics[width=\linewidth]{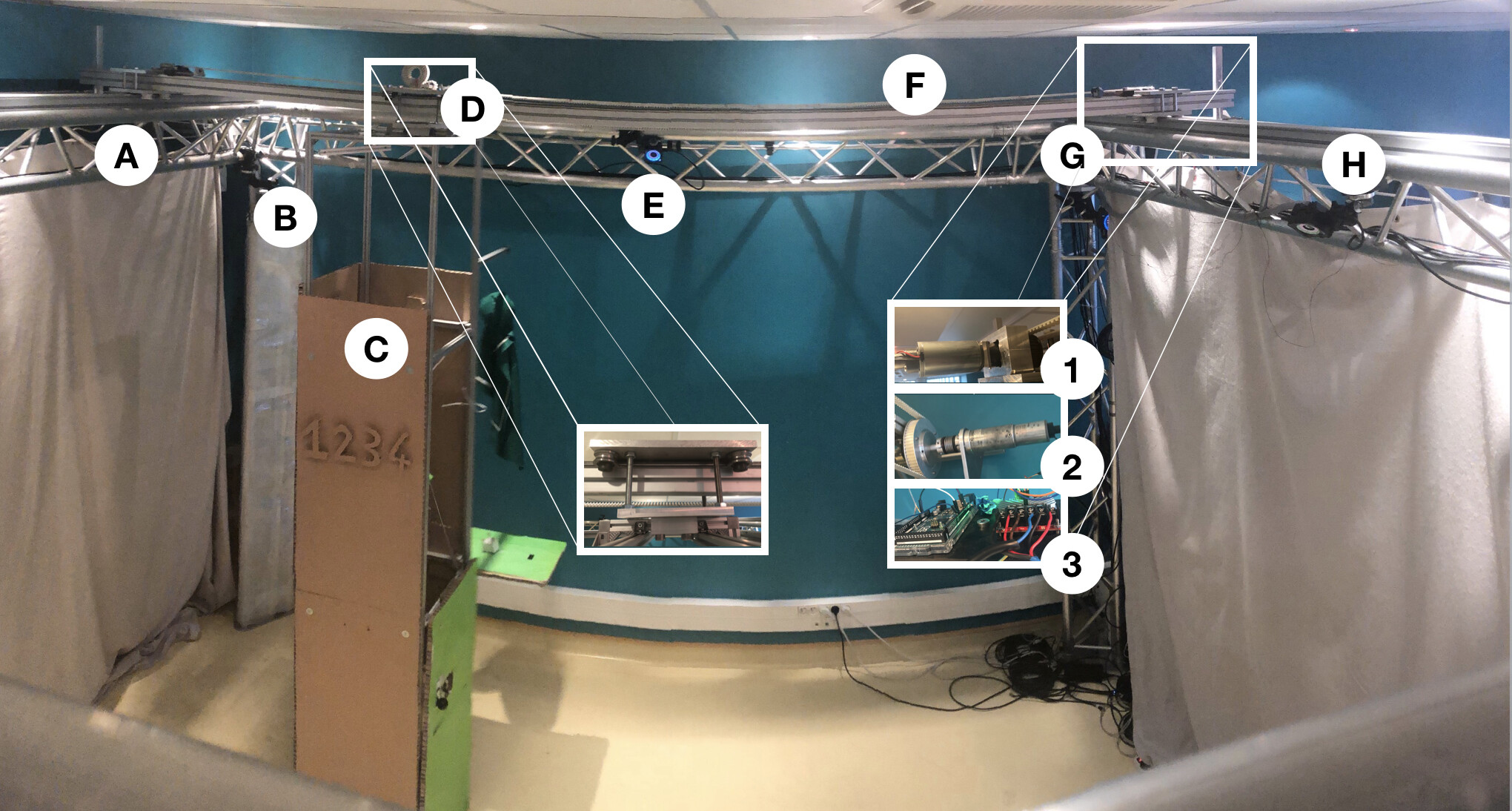}}{A detailed picture of CoVR with an isometric view. The gantry supporting CoVR (A) is visible, as well as its rails. Call-outs show CoVR’s motors, pulley-belt mechanisms and electronics.}
  \caption{Top isometric view of \SYSTEM/ setup: (A) Structure; (B) Skeleton, modular column-like structure to attach props and panels; (C) \SYSTEM/ panel; (D) Carriage; (E) $X$-axis rail; (F) 1: $X$-Pulley-belt system and Motor, 1: $Y$-Pulley-belt system and Motor, 3 - Electronics (Arduino and RoboClaw); (G) $Y$-direction rail.}
  \label{fig:setup}
\end{figure}

When designing our \textit{Robotic Shape Display}, \SYSTEM/, we primarily focused on \textbf{force feedback} and \textbf{workspace size} as design considerations (which is a challenging trade-off for haptic devices \cite{seifi_haptipedia:_2019}) as well as speed, interaction opportunities, price and safety. We initially considered mobile robots such as \cite{he_robotic_2017}. However, these interfaces are limited by a compromise between force-feedback, speed and autonomy. More particularly, we decided to focus on enabling strong force feedback at a whole-body scale and allowing different body postures.
We then deliberated upon a \textit{grounded} solution, a 2D Cartesian ceiling-mounted robot, which can be integrated into a room-scale arena\footnote{The robot can be mounted on the ceiling or on an external truss structure (\textit{triangle aluminum Global Truss}) as shown in Figure \ref{fig:setup}.} (4x4x2.5m; \textit{LxWxH}).  
The advantages are speed, accuracy, force-feedback, while allowing to move potentially heavy physical objects without the embedded mass affecting its displacements. 

Another important consideration was the number of degrees of freedom (DoF) of the robot ($X-Y$ planar motion, $Z$ elevation, $W$ rotation around $Z$, 6DoF Robotic arm...). When dealing with robotic interfaces, a trade-off between price, complexity and interaction possibilities can be drawn. We realised that 2D planar motion carrying a modular structure already allows quite a large variety of scenarios while keeping a low technical complexity and cost. However, our chosen architecture can be extended with additional DoFs (for instance, by attaching a 6DoF Kuka robotic arm; see section Discussion).

In the following sections, we describe the main components of the final\footnote{We used an iterative process to design \SYSTEM/. At each iteration, we improved key features such as robustness, accuracy, speed, safety, while widening the interaction possibilities.} version of our system. We then evaluate \SYSTEM/ in a technical evaluation validating its control through a user-intention based algorithm.




\subsection{Robotic system}
\textit{Robot}. \SYSTEM/ relies on a 2D Cartesian ceiling robot (Figure \ref{fig:setup}), actuated with DC Motors (\textit{Dunkermotoren 55x30, KPL43 gearbox,$1.81 Nm$ torque for X-axis, Dunkermotoren 63x55, KPL57 gearbox, $9.75 Nm$ torque for Y-axis}) trough a pulley-belt mechanism (Figure \ref{fig:setup}). We chose a pulley-belt mechanism because it is simple to implement and can easily be scaled to larger VR arena. The robot moves a $15x15cm^2$ carriage on which is attached a modular structure (see below). 

The robot is controlled in speed with a \textit{Roboclaw 2x30A V5E} motor controller and \textit{AEAT-601B-F06} encoders, mounted on a custom-designed 3D-printed support. The Roboclaw controller is connected to an \textit{Arduino MEGA 2560} micro-controller. It provides closed loop control with a \textit{PIV}\footnote{Proportional position loop Integral and proportional Velocity loop} scheme. The total price of the robot (motors, rails and pulley-belt) is under 1500euros.


\textit{Speed}. The speed of the robot depends on the distance to travel. For large distances (> 1 m), the speed is over \textbf{1.1 m/s}, which is approximately a normal human walk speed. 
For small distances (< 80 cm), speed is about 0.5 m/s, which remains faster than current mobile solutions (e.g. \cite{suzuki_roomshift_2020, he_physhare:_2017}).  



\textit{Noise.} At full speed, \SYSTEM/'s average noise is 55 dB average (max: 65 dB).

\textit{Weights and Forces Capabilities.}
The carriage can support a total weight of $800N$ vertically ($\approx 80kg$) and $1000N$ horizontally. The embedded mass the carriage can support is large enough to support a human lying on it or even to be pushed by it (Figure \ref{fig:top_photo}-A,B) without causing any damage to the structure. The system can also provide high traction force to pull the user (-C) or even transport her (-D).

\subsection{Column}
\begin{figure}[h]
  \centering
\pdftooltip{  \includegraphics[width=0.8\linewidth]{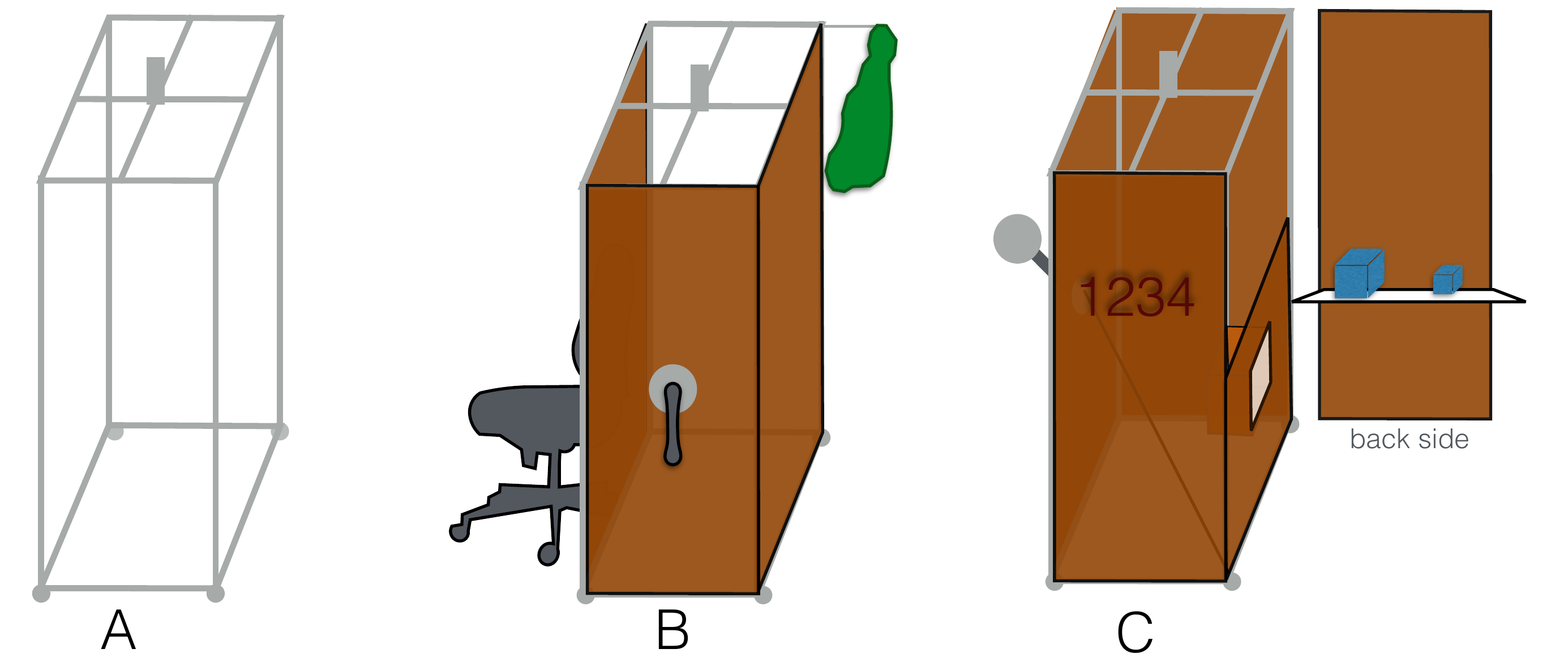}}{Schematic/Drawing of different column configurations. The skeleton by itself is drawn in (A), while (B) and (C) reflect some scenarios configurations. }
  \caption{Column design. (A) Modular structure attached to the 2D ceiling robot to provide a wide variety of surfaces and props. (B) The 3-side column used in the user study with a chair (left), a cylinder attached to a spring virtually representing a broom (front), a large cardboard simulating a wall and piece of fabric representing a ghost (right). (C) A 4-side column implemented with a lever attached to the structure with an elastic (left), haptic code made in cardboard and glue (front), and a tray with a large and small cube (back) to insert into the locker (right).}
  \label{fig:column}
\end{figure}

A column-like modular structure (Figure \ref{fig:column}-A) is attached to the moving carriage. Different surfaces in arbitrary positions, shapes, orientations\footnote{This feature made the $W$ rotation partially redundant.} and sizes can be attached using a simple clamping mechanism. It is similar to stage designing in real theatres \cite{pair_flatworld:_2003}, where a limited number of decors can quickly be replaced. Another advantage is to easily support \textit{DIY}: the stage designer can use cardboard or props with different mass, textures and shapes that users can freely manipulate at different heights of the column. The positions and shapes of the physical objects are then communicated to the VR designer in a calibration phase.
In summary, the column has been designed to be flexible enough to support a wide range of interactions. Figure \ref{fig:column}-B and -C show two examples of implemented columns. The section Interactions and Demo Applications detail interactions with these columns.

\subsection{Display and Tracking} We use the Oculus Rift S \cite{noauthor_oculus_2019} HMD because it is not sensitive to occlusion problems and it allows interactions under or even in the column (Figure \ref{fig:postures}). We used Unity3D to create virtual scenes. It centralises the communication and synchronisation between different components though plugins (SteamVR, Arduino/Roboclaw). In particular, the SteamVR plugin asset \cite{noauthor_steamvr_2019} is used for the Oculus communication and the Uduino package \cite{teyssier_uduino_2019} for fast prototyping between Arduino and Unity. 



\subsection{Safety}
As users are invited to move around an active large-scale mechanical system, safety measures had to be established. These were planned on several levels, from the structure conception to the motions around the users during interactions.

\textit{Carriage:} One risk is the fall of the carriage. The carriage can support both larger axial (800N) and radial (1000N) forces than those required for the envisioned scenarios. A security coefficient of 2.5 was introduced for elastic deformation calculations in the conception process.

\textit{Column motions:} Hardware, software and electronic emergency stops are implemented. The carriage motion is restricted on both ends with spring-based mechanical stops. 
The software stops the motors when the column is within $2cm$ of these limits. The controller electronically shuts down when the motor's current exceeds $5A$. More importantly, the column immediately stops if the user is not tracked for more than 0.5s. Finally, the \textit{game master} has a manual emergency stop button that turns the system off, keeping it electrically grounded to avoid potential shocks. Finally, given the power and speed of the robot, it is important to ensure the column will not accidentally physically collide with the user. We thus developed an algorithm to generate the robot trajectories.



\subsection{Robot Motion Control}
We present a model to control the robot displacements\footnote{The algorithms, models, scripts and user data are available on this repository: https://www.bouzbib.com/CoVR/ 
}. While trajectories are easily generated by the Cartesian structure (XY displacements), the algorithm inputs for scenarios involving multiple objects of interest need to be defined and safety measures around the user need to be implemented. 
The main idea is to attach the robot to a virtual proxy (a ball with mass and gravity) with a spring-damper model.
The ball's displacements depend on (1) the user's location to avoid collisions, (2) the user intentions and (3) the progress of the scenario to attract the ball towards the objects users are most likely to interact with next. A key contribution regarding our trajectory generation model is the elaboration of a low-computational \textbf{user intention model} working with common HMDs. We now detail our approach to generate trajectories.






\textit{Trajectory generation.} Each virtual object of interest $i$ within the scene gets a weight $W_i$ which depends on its likeliness to be interacted with next. The virtual proxy (ball) and CoVR command position $\SYSTEM/(x, y)$ is hence a weighted average of the positions of each object of interest:
\begin{eqnarray} \label{eq:position}
    \SYSTEM/(x, y) =  \frac{\sum_{i = 1}^{N} W_i * (x_i, y_i)}{\sum_{i = 1}^{N} W_i}
\end{eqnarray}
where $N$ is the number of virtual objects of interest (VOI) in the scene, ($x_i$, $y_i$) the cartesian coordinates of the VOI $i$ and $W_i$ its weight, estimated given a user intention model (see below). A virtual spring between the proxy and the command position is then defined, and the according spring force is applied for the proxy to reach this position. We use Unity3D's physics engine to automatically generate the proxy trajectories to reach a target. The target position is not necessarily the position of a virtual object. 
If the scene contains two objects of same interests, \SYSTEM/ will automatically place itself between these two objects' positions, hence the displacement when one becomes the chosen object of interest is minimised and \SYSTEM/ is more likely to reach it prior to user interaction. As the proxy is also attached to the \SYSTEM/, its resulting motion takes naturally into account the robots speed limitations.

\vspace{-0.3cm}
\begin{figure}[h]
  \centering
  \pdftooltip{\includegraphics[width=\linewidth]{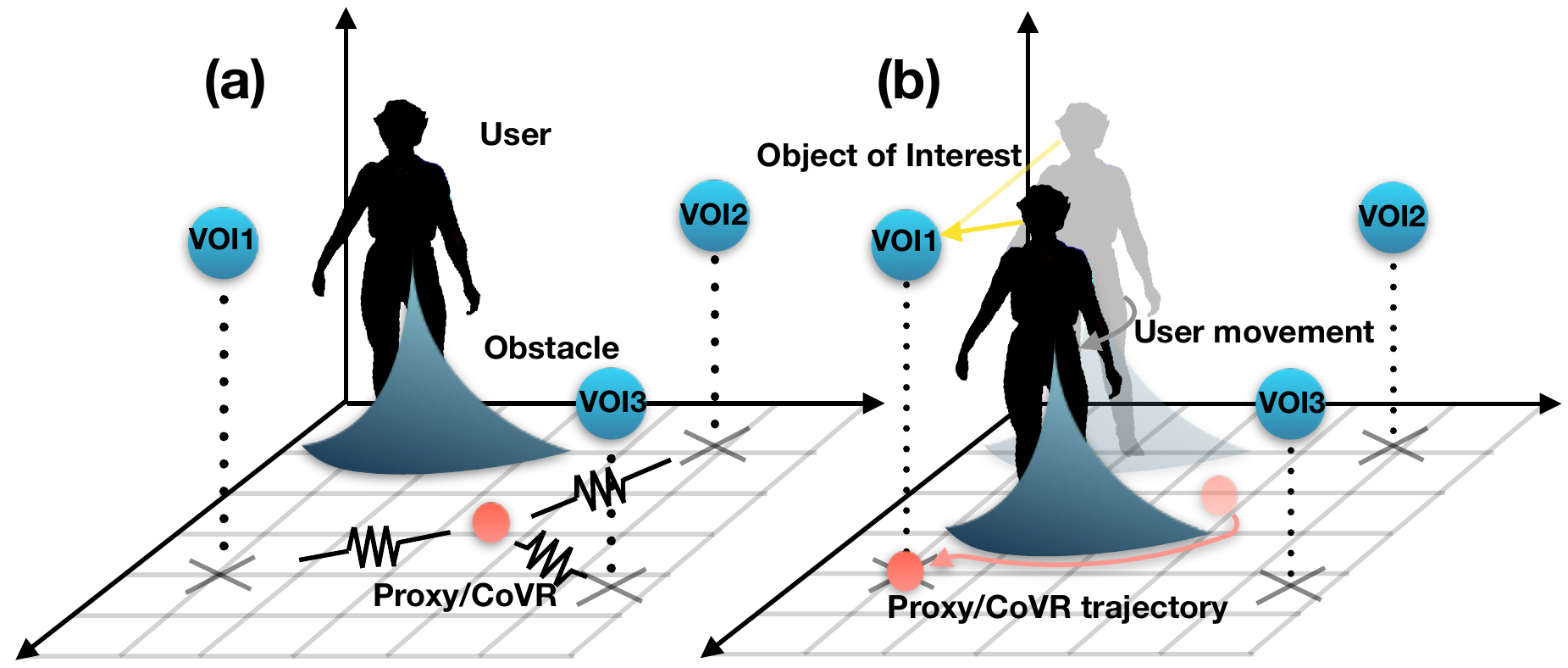}}{(a) A 3D referential is drawn with a grid. A virtual avatar is at the back of the grid, and a cone-like shaped obstacle is around the user. Three virtual objects of interest are available. A ball is representing the proxy/CoVR, and a spring is attached from this ball to each obstacle. (b) A yellow raycast shows the user is choosing one of the VOI. The avatar has moved (and its previous position is shadowed). The ball trajectory is drawn: it is avoiding the user obstacle by going around it and reaches the VOI’s position.}
  \caption{Control algorithm relying on a physical model: (a) The virtual proxy of the physical \SYSTEM/ column is connected to all virtual objects of interest (VOIs) with weights depending on the users' intentions to interact with them. The user and other forbidden zones are covered by a rigid cone-like obstacle to be repulsive. (b) Whenever the user is about to interact with a VOI, the proxy/\SYSTEM/ move towards it, while naturally avoiding obstacles (e.g the user).}
  \label{fig:algo}
\end{figure}
\vspace{-0.3cm}
We also created virtual obstacles to cover all forbidden areas in the arena (user, external people, furniture). Indeed, we designed a cone-like rigid shape that we attached to the user's position as shown in Figure \ref{fig:algo}. The size of the cone (diameter = 90cm) was chosen to avoid collisions even if the users' arms are open. Thanks to the contact mechanics and gravity in Unity, the proxy is naturally pushed and rolls away from the obstacle. The obstacle's curvature ensures a smooth deceleration of the proxy when this latter is getting to close to it. The radius of the obstacle decreases (20cm) when the user comes near a object of interest, so the proxy is not pushed away. 


\vspace{-0.3cm}
\subsection{User-intention model}
We elaborated a user intention model to support non-deterministic scenarios, i.e. scenarios where the system does not know beforehand which object to physically overlay.
The model inputs are the positions of the virtual objects of interest (VOI) as well as the available data from the users' apparatus: the HMD's position and orientation. It hence does not require additional hardware such as eye-tracker or finger/hand tracker.

We defined the total weight $W_i$ of the VOI $i$ to be a function of the user's distance $D$ to a VOI, and her orientation ($O$). 
\begin{eqnarray}
    W_i(d, \theta)  &=& \omega*D(d) + (1-\omega) *O(\theta) \\
    W_i(d, \theta) &=& \omega* \frac{1}{1+d} + (1 - \omega) * e^{(cos(\theta) - 1)}
\end{eqnarray}
where $\omega$ is the contribution of the distance over the orientation. $D(d)$ and $O(\theta)$'s ranges are between 0 and 1, hence $W_i$'s range is from 0 to 1 too. 
$O(\theta)$ is equal to 1 whenever the user's HMD orientation is colliding with any surface point of the VOI's mesh, and is decreasing exponentially whenever the user's orientation moves further away. On the same principle, $D(d)$ is equal to 1 whenever the user is close to a target, and decreases with the same regards\footnote{We also increase the stability of the column in the vicinity of the VOI. When $W_i > 0.8$, $W_i$ is rounded to 1, typically when an object is at less than 20 degrees from the user's HMD direction or when the object is at a distance below 20cm from the user. It allows for \SYSTEM/ to stay at the closest VOI as long as the user remains in its vicinity.}.

\subsection{Scenario-based model}

Depending on the progress of their scenario, designers can estimate the prior probability of an object to be interacted with: in a basketball game for instance, the user is more likely to interact with the ball first than with the hoop. We let the possibility to designers to define their own scenario-based model by refining the estimation of $W_i$:
\begin{eqnarray} \label{eq:position}
    W_i = P_i \times W_i(d,\theta)
\end{eqnarray}
where $P_i$ is the prior probability of the VOI $i$ to be interacted with from the progress of the given scenario.
We will discuss the use of these probabilities in the Discussion section of our Technical Evaluation below.

\section{Technical Evaluation}


The primary aim of this technical evaluation is to determine the $\omega$ parameter of the user intention model, i.e. the optimal contribution of the distance over the orientation to estimate which object of interest is more likely to be interacted with.
We are also interested in studying \SYSTEM/'s success rate as a function of the number of objects of interest (\textit{distractors}) within the scene. Indeed, we anticipated that the performance of the user intention model and the value of $\omega$ depend on the number of distractors within the scene.
Finally, we want to confirm that \SYSTEM/'s speed is sufficient enough to reach a virtual object of interest even when the user does not have a decision to make (number of VOI = 1).



We first perform a data collection over a panel of users to better understand how intentions can be quantified as a function of both distance and orientation. We then perform multiple physical simulations to find the best $\omega$ parameter that matches users' behaviors.



\subsection{Data Collection}

\textit{Participants and Apparatus.} 6 participants (3 male, 1 left-handed) aged from 26 to 32 (average = 28; std = 2.0) volunteered for this experiment. All participants were familiar with VR and were asked to wear the Oculus Rift S. Users also wore Optitrack markers on their dominant hand. The Oculus headset was also equipped with Optitrack markers, for an accurate tracking in space. The virtual scene was created using Unity3D game engine.


\subsubsection{Experimental Design}

\begin{figure}[h]
  \centering
  \pdftooltip{\includegraphics[width=\linewidth]{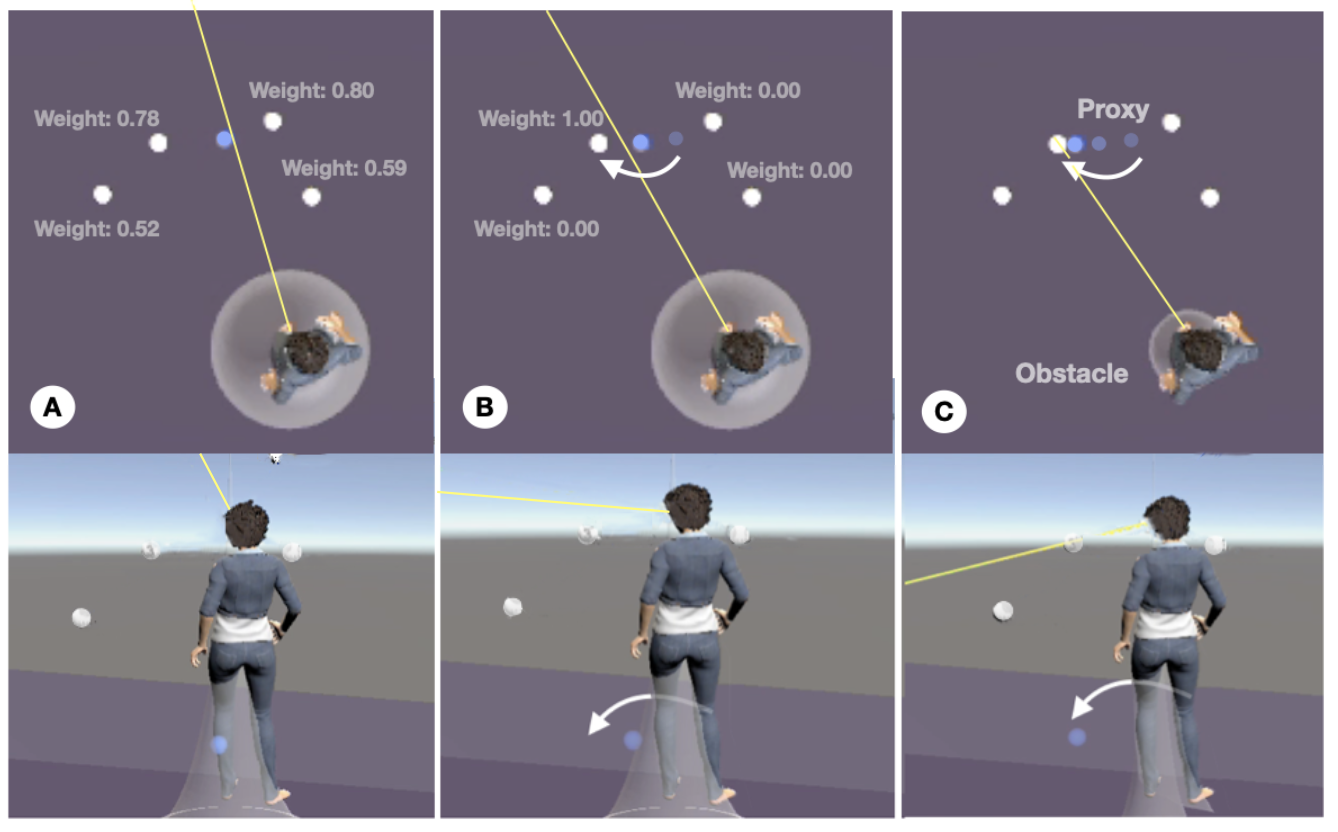}}{Top views and Isometric views of the technical evaluation 3D scenes. The target and three distractors are available in the scene (illustrated as white balls), and their weights are written next to them. A yellow raycast displays the avatar’s orientation. The proxy is illustrated by a blue ball. (A), (B), (C) show the user picking a VOI, and its weight goes to 1. In the meantime, an arrow shows the proxy’s displacement towards this VOI. Whenever the proxy is in place, the obstacle around the user decreases in size.}
  \caption{Technical Evaluation \textit{"Simulation"} Virtual Scene example after the Data Collection. (A) User looks for the target (according to the walls' instruction). Weights change according to her position and orientation. (B) \textit{Intention Detection}: User chooses a target and its weight goes to 1. (C) \textit{Trajectory}: The proxy (blue ball) moves accordingly with the centroid of all the objects' of interest's weights towards the chosen one (weight = 1), while avoiding the user obstacle. When the proxy reaches the chosen ball, the user obstacle size decreases.}
  \label{fig:tech-eval-image}
\end{figure}

\textit{Task and Stimuli.} We considered an exploratory task, such as the ones users would perform in games, i.e. users take their time, observe the decors, avoid virtual obstacles and face their objects of interest whenever interacting. 
To replicate these game features and to capture the corresponding users' behaviors, we created an empty scene where virtual numbered balls appear simultaneously at random locations with random orientations (see Figure \ref{fig:tech-eval-image}).
Instructions are written on the walls surrounding the users, and tell them to touch a given numbered virtual target. Users are then asked to face the targets whenever touching them.

\textit{Conditions.} In this experiment, we control the number of distractors within the scene from 0 to 4 (number of balls is from 1 to 5). This allows us to understand the performance of \SYSTEM/ over the number of available VOIs. The minimum distance between two targets is their diameter - 10cm (eg they cannot overlap) and they cannot appear at the user's location. As long as the user does not touch the target ball, nothing changes in the scene. As soon as the target ball is interacted with, another condition starts. 

\textit{Design.} 
We used a within design. All participants tested all five conditions (0,1,2,3,4 distractors). The order of appearance of each condition was randomized within the blocks. Participants performed 10 blocks. The duration of the experiment was about 12 minutes per participant (std = 2.6). In summary, the experimental design is: 6 participants $\times$ 10 blocks $\times$ 5 conditions $=$ 300 trials. 


For each trial, we measure the users' position and orientation at each frame, with a frame rate of 75 fps.

\vspace{-0.3cm}
\subsection{Parameter Fitting}

We used the data collection to replicate the users' displacements into a Simulation virtual scene. The robotic system physically moved accordingly with our "user intention model" (section above). Each simulation corresponded to a different $\omega$. We simulated all the data from the 6 participants (i.e. including the 5 conditions). We first performed a broad exploration of $\omega$ (step= 0.25) and then refined it to find the optimal one for each condition (number of distractors in the scene). We tested 13 parameters over 6 users, which resulted in more than 17 hours of simulation.

Our main measurement was the success rate of \SYSTEM/ reaching a VOI before the user, i.e. when \SYSTEM/'s distance to the target was below its diameter (10cm) when the user was touching it. 

\subsection{Results}

\begin{figure}[h]
  \centering
  \includegraphics[width=\linewidth]{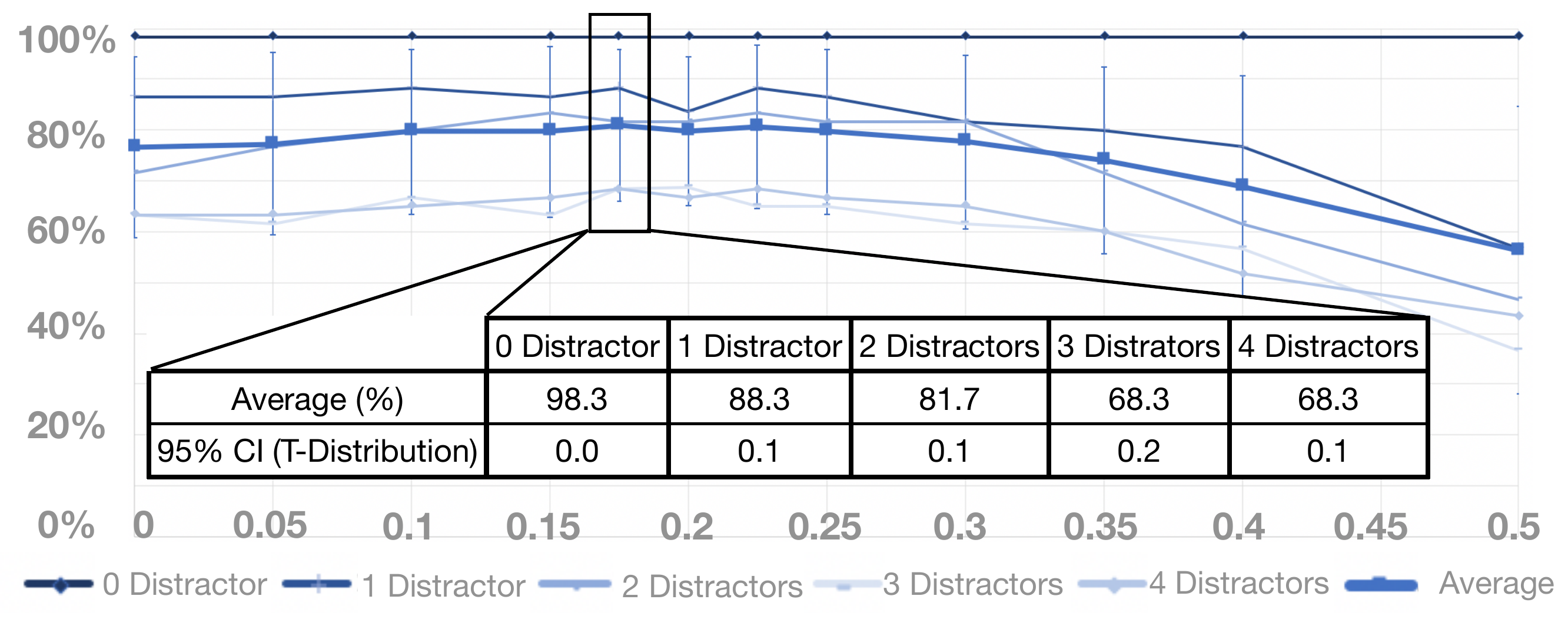}
  \caption{Success Rate of \SYSTEM/ reaching the chosen VOI prior to the user interaction, function of $\omega$ and the number of distractors. Error bars indicate 95\% confidence interval with a T-Distribution. The table shows the Success Rate with the optimal parameter, $\omega$ = 0.175, as function of the number of distractors.}
  \label{fig:tech-eval-results}
\end{figure}

\textit{Success Rate}. Figure \ref{fig:tech-eval-results} shows the success rate as a function of $\omega$ and the number of distractors. The success rate is approximately 100\% (only 1/300 targets missed) when there is only one VOI in the scene, indicating that the system is at least as fast as the participant when the target position is known (i.e. the system does not rely on the users' intention). The results also confirmed that the success rate decreases with the number of distractors. Figure 6 also shows that we obtain the best average success rate (80\%) with $\omega$ = 0.175 (CI=14\%) regardless of the number of distractors. Success rate remains above 80\% up to 2 distractors.

We also note that the success rate per user decreased with the time spent in the experiment (88\% success for a 14mn experiment vs 74\% for an 8mn experiment). 




\textit{Target distance}. We measured the average distance between the carriage and the target centres when the user was colliding with the virtual target. The average distance among all the trials is 1.8cm (95\% CI = 0.33 cm) demonstrating the repeatability of our implementation.


\textit{Detection time}. We also measured the time difference between the target's weight reaching 1 and the user colliding with it. Results show that this detection time does not depend on the distractors, with a 7s average (std = 0.6s) and a 96\% accuracy. We note that if the detection time is below 4s, it results in a failure of the overlaying, as \SYSTEM/ struggles to get around the user (especially the obstacle) and place itself properly.

\textit{Number of users collision}. No collision between the user and \SYSTEM/ were noted during the simulations.

\textit{Accuracy}. Finally, we measured the distance between the virtual proxy and the physical column. The mean distance over all users and conditions is 0.94 cm (CI 95\% = 0.99 cm), which ensures they share the same trajectory, and hence a safe user environment around \SYSTEM/.

\begin{table}[h]
  \centering
  \includegraphics[width=\linewidth]{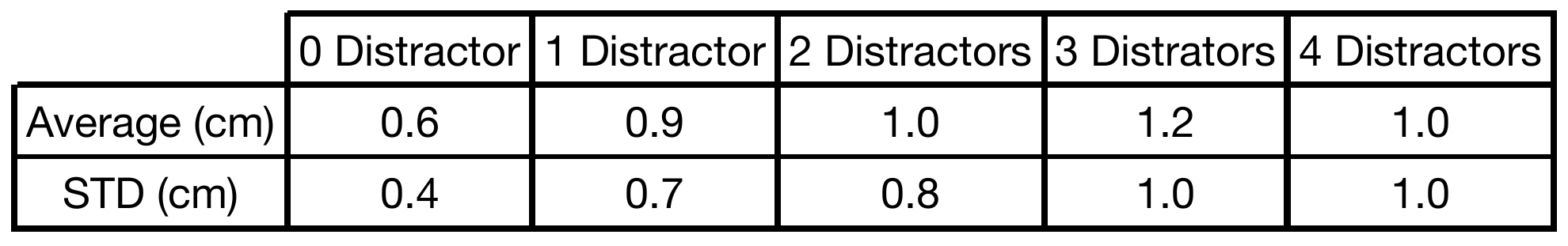}
  \caption{Accuracy, measured by the distance between \SYSTEM/ and the proxy, with $\omega$ = 0.175.}
  \label{fig:table_accuracy}
\end{table}
\vspace{-0.3cm}
\subsection{Discussion}

This evaluation tested \SYSTEM/ in an uncontrolled environment, with \textbf{random} locations and orientations for each target and distractor and a user-intention based model. Despite this environment, our system had a high success rate ($>80\%$) with three virtual objects of interest while preserving the user's safety (no collision). Multiple directions can be envisioned to increase this success rate in non-deterministic scenarios.

\textit{Adding a scenario-based model.} According to the Equation \ref{eq:position}, we can add "prior probabilities" to the different VOIs, depending on the progress of the scenario. After selecting our optimal parameter $\omega = 0.175$, we ran the robotic simulation by adding to the actual target a 75\% probability to be interacted with. The distractors were hence splitting the remaining 25\% of interaction probability. The results are summarized in the Table \ref{fig:table_probs} and confirm that adding a scenario-based model improves the prediction with a success rate higher than 93\% even with four distractors.



\begin{table}[h]
  \centering
  \includegraphics[width=\linewidth]{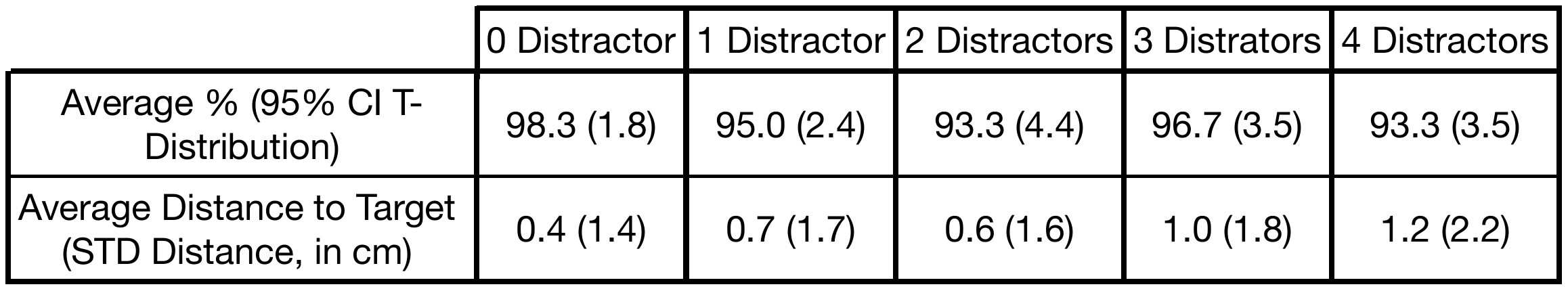}
  \caption{Success Rate and Distance to Target with $\omega$ = 0.175, and a 75\% probability to be interacted with added on the target.}
  \label{fig:table_probs}
\end{table}


\textit{Assigning multiple VOI to the same Physical position.} 
Thanks to its size and shape, \SYSTEM/ can contain multiple objects on different panels and at different heights. We can take advantage of this feature to assign multiple virtual objects of interest to the same physical location, hence reducing the amount of \SYSTEM/ displacements and the risk of spatial mismatches.




\textit{Adding visual effects}. When a spatial mismatch is likely to occur, literature usually proposes to cater for it with visual effects \cite{cheng_turkdeck:_2015} or dynamic redirection techniques \cite{abtahi_beyond_2019-1, razzaque_eurographics_2001, azmandian_haptic_2016-1}. These respectively distract the users and give spare time to the robot to reach the target location or dynamically correct the users and \SYSTEM/'s interaction positions.

\section{Interacting with \SYSTEM/}







The technical evaluation showed that \SYSTEM/ is able to move at a sufficient speed to follow users at a natural walk speed. 

When \textit{no interaction is required}, \SYSTEM/ remains out of reach and the users can wander in the whole arena. \SYSTEM/ thus does not interfere with users' natural behavior. Letting the users truly walk (instead of using a metaphor for locomotion) reinforces the immersion \cite{usoh_walking_1999}.

When \textit{interactions are required}, a key aspect of \SYSTEM/ is to allow interactions involving strong kinesthetic feedback at a body scale. We distinguish two main uses of \SYSTEM/: \textbf{static} use where users transmit forces when interacting with the column (e.g. exploration, manipulation) and \textbf{dynamic} use, where users are receiving forces enabled by \SYSTEM/'s displacements during the interaction (e.g. leading through forces, transport). We now detail these two uses of \SYSTEM/.

\subsection{Static Use of \SYSTEM/}

\begin{figure}[h]
    \centering
    \pdftooltip{\includegraphics[width=\linewidth]{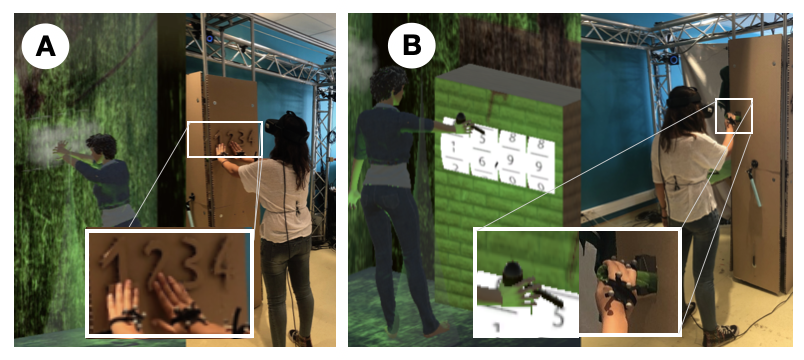}}{(A) and (B) show virtual and physical correspondences. (A) illustrates the user exploring a code (hidden in VR, and in relief in the real world). A call-out shows the hand exploration and the code : 1234. (B) shows the user entering this code in a virtual slot machine. A call-out shows the virtual lever and the physical one - a cylinder coming out of CoVR’s panel, attached to CoVR’s skeleton with an elastic to allow for the directed manipulation.}
    \caption{(A) Tactile Exploration: The user tactilely explores large surfaces, for instance, to find a hidden code over a human-sized wall. (B) Directed Manipulation: The user pulls a lever which is attached to CoVR with an elastic, letting it a single degree of freedom, providing a mechanical manipulation of props.}
    \label{fig:interactionBloc}
\end{figure}

\textit{Hand exploration:} Hands remain the primary body part for exploring the world and the most sensitive one.
Users can probe objects directly with their bare-hands. As such, interactions are not limited to one finger: surfaces can be realistically touched and their texture fully felt with the whole hand. In particular, users can interact with the palm, which contributes to a sense of tangible presence, as it enables kinesthesia on top of tactile cues \cite{minamizawa_simplified_2010}. Moreover, the explored surface can be large and not limited to a specific orientation or shape. For instance, users can perform large hand movements to find a specific tactile pattern on a wall for instance (Figure~\ref{fig:interactionBloc}-A). 

\textit{Whole-body interactions:}
Users can apply strong forces with any part of their body: users can lean on a fixed wall (Figure \ref{fig:top_photo} - B), push hard on it with their hands or shoulders (Figure \ref{fig:Level1}) or even kick it. \SYSTEM/ is rigid and robust enough to remain still during all of these interactions.


\begin{figure}[h]
    \centering
    \pdftooltip{\includegraphics[width=\linewidth]{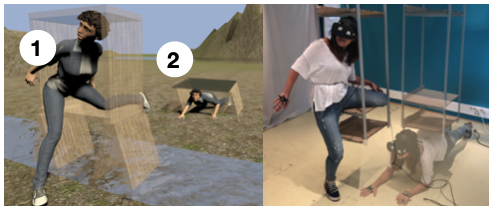}}{Virtual and physical correspondence of a obstacle course. The avatar is (1) going through an obstacle, and crawling under another one. The user is going through CoVR’s skeleton in an XY-position and crawling under it in another position.}
    \caption{Postures; The user 
    (1) goes through an obstacle with constrains below and beneath her (2) or crouches.}
    \label{fig:postures}
\end{figure}

\textit{Postures:} \SYSTEM/ also supports a variety of users' postures with interactions at different heights such as crouching under a table, going through obstacles with physical props both below and above the users (Figure \ref{fig:postures}), sitting on a chair \cite{zielasko_either_2020, teng_aarnio_2019} or climbing a stair to reach a high target (Figure \ref{fig:Level1}).

\begin{figure}[h]
    \centering
    \pdftooltip{\includegraphics[width=\linewidth]{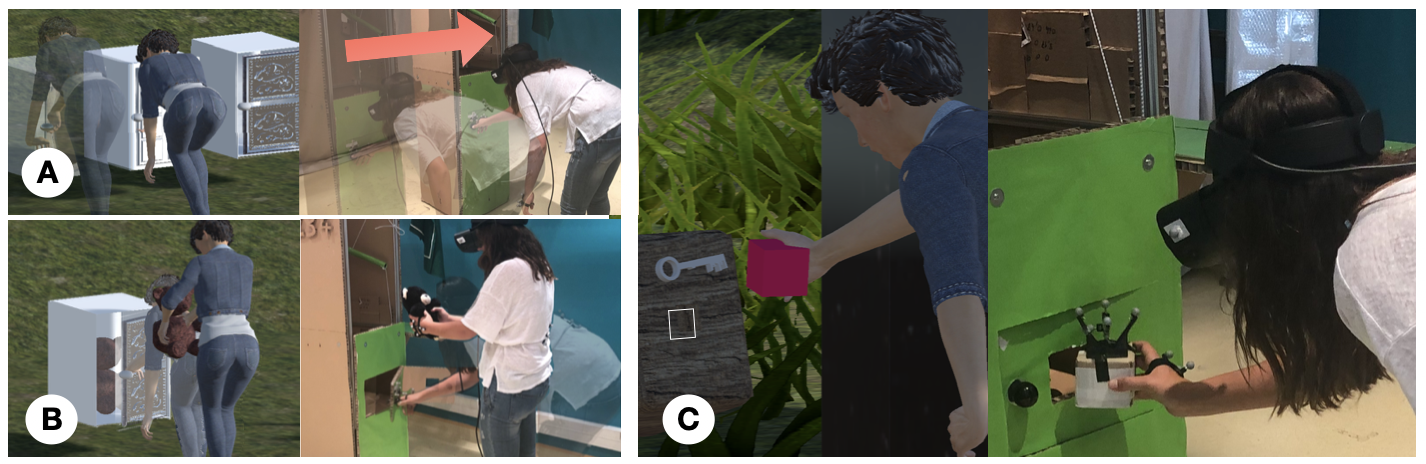}}{Three virtual/physical pictures. (A) The avatar opens three different doors. In the real world picture, a red arrow shows CoVR’s displacement, and the user opening multiple physical doors (a shadow effect is used to illustrate the different positions of both CoVR and the user). (B) The avatar finds a teddy bear and manipulates it. The user does the exact same thing. (C) The avatar is holding a cube and putting it in a virtual locker. In the real world, the user hold a real cube and tries to fit it in a cube-shape slot, which is too small to fit the actual cube.}
    \caption{(A) Directed manipulation; User opens three virtual doors - but only a single physical one, cut through a panel cardboard. (B) Free manipulation; User finds a teddy bear. (C) Free manipulation and Contact; The user manipulates a cube which is too big to fit in the locker. She realises she needs to find a smaller cube.}
    \label{fig:manip}
\end{figure}
\begin{figure}[h]
    \centering
    \includegraphics[width=0.9\linewidth]{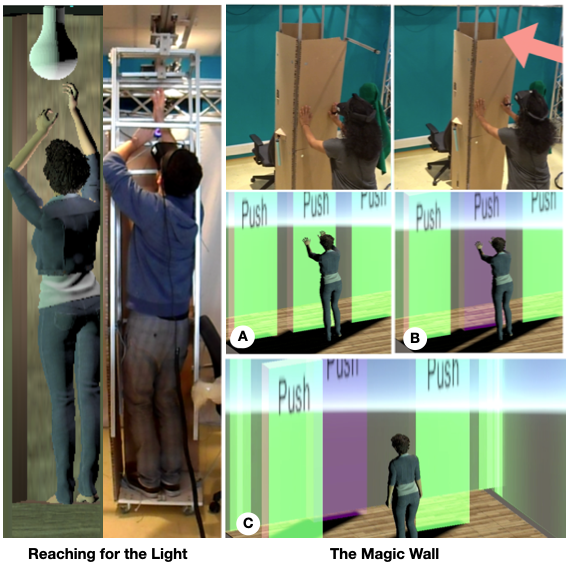}
    \caption{"Escaping the Room". \textit{Reaching for the Light}: User climbs in the cupboard to reach the virtual light bulb; \textit{The Magic Wall}: (A) User chooses a wall. \SYSTEM/ moves accordingly with the user's intentions. User pushes on the wall. \textit{We note that none of the users touched the "ghost" by accident during the experiment.} (B) The wall remains static, and changes color to encourage the user to maintain contact. (C) After 10seconds of maintained contact, the Magic Wall moves, giving the user the impression of having pushed it herself.}
    \label{fig:Level1}
\end{figure}

\textit{Manipulation:} Manipulation of real objects and passive props improves interaction fidelity \cite{rogers_exploring_2019, insko_passive_2001}. \SYSTEM/ enables different types of object manipulation:
\begin{itemize}
    \itemsep0em 
    \item \textit{Free manipulation}. \SYSTEM/ can carry untethered objects which users can grab and freely manipulate. A large variety of samples (Figure \ref{fig:manip}-B) of any textures is possible, as long as dimensions and weights are compatible. Thanks to the \SYSTEM/'s grounding and high motor torques, it can carry large masses without compromising its speed or accuracy.
    \item \textit{Contact.} Objects can also be manipulated to interact with each other. For instance, in Figure \ref{fig:manip}-C, the big cube does not physically fit in the locker. The user hence needs to find a smaller one. 
    \item \textit{Directed manipulation}. Users can interact with objects tethered to \SYSTEM/. Its structure allows for mechanical manipulation of objects and for users to actuate them. For instance, in Figure \ref{fig:interactionBloc}-B, the user actuates a lever mounted on the column, simulating a slot machine. By attaching objects on \SYSTEM/'s skeleton, mechanical manipulation with multiple numbers of degrees of freedom is possible.
\end{itemize}

A single physical object can overlay multiple virtual ones of similar primitives \cite{hettiarachchi_annexing_2016}. Instead of using visual effects such as \cite{azmandian_haptic_2016}, \SYSTEM/ physically moves a single prop to overlay multiple virtual ones. For instance, one physical door can overlay three virtual ones (Figure \ref{fig:manip}-A).
These mappings were previously seen in the literature \cite{he_physhare:_2017, he_robotic_2017}. 


\subsection{Dynamic Use of \SYSTEM/}

In the previous section, \SYSTEM/ was motionless during the interaction (\textbf{static}). The following interactions require the system to move in the user's vicinity (\textbf{dynamic}). 



\textit{Receiving Physical Contact:} \SYSTEM/ can physically touch the users and produce impact force feedback \cite{wang_movevr_2020}. It is thus initiating the haptic interaction, instead of the user. As receiving an interaction might be surprising in VR, we recommend attaching props at a distance from \SYSTEM/'s main skeleton, to produce light impact forces.
For instance, a fabric (60cm away from the main skeleton) can lightly brush the users to simulate the crossing of a ghost (Figure \ref{fig:ghosts}) through them. Users can also be touched by a virtual agent trying to catch their attention, providing a sense of physical presence \cite{lepecq_afforded_2008, hoppe_human_2020}.


\textit{Leading through forces:} Users can be led by \SYSTEM/ through body-scaled tension and traction forces. For instance, in Figure \ref{fig:top_photo}-C, the user physically holds a cylinder attached by a spring to the column, virtually represented by a broom, which provides her with a large force-feedback and leads her the way in the virtual environment. She is pulled by the broom when \SYSTEM/ moves. Another example is inspired from \cite{cheng_mutual_2017}, involving a fishing pole where the line is attached to the column. The motion of the column creates the illusion of a fish biting. 

\begin{figure}[h]
    \centering
    \includegraphics[width=0.9\linewidth]{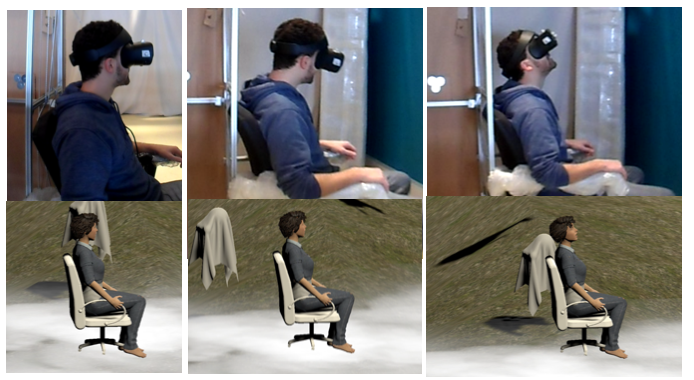}
    \caption{\textit{Moving in the Clouds}: User is sitting in a chair and physically transported through the clouds.}
    \label{fig:travel_clouds}
\end{figure}

\begin{figure}[h]
    \centering
    \includegraphics[width=0.8\linewidth]{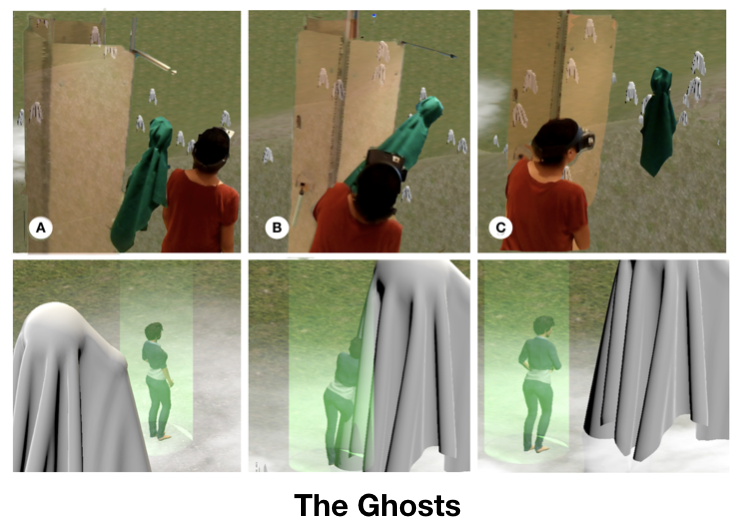}
    \caption{\textit{The Ghosts}: (A) User enters the halo. (B) A piece of fabric lightly brushes the user's head. (C) The ghost flies away.}
    \label{fig:ghosts}
\end{figure}

\textit{Transport:} 
Finally, \SYSTEM/ mechanical properties open up a new range of interactions in VR. Indeed, \SYSTEM/ can transport the users. For instance, it can move a chair with a sitting user to a different location (Figure \ref{fig:travel_clouds}) as \SYSTEM/ can handle large embedded masses. We envision other scenarios transporting the user, such as Wind-surfing or Water-skiing \cite{ye_pull-ups:_2019}.



\section{Demo Applications}
We created a two-scene demo application to demonstrate the interaction possibilities offered by \SYSTEM/. It relied on the 3-side column illustrated in Figure \ref{fig:column}-B and involved 5 interactions, 7 virtual objects but only 5 props. 
In the following subsections, user interactions will be displayed in \textbf{bold} while the motions and \SYSTEM/'s interactions will be displayed in \textit{italics}.

\subsection{Escaping the Room}

We created a first scene where the users need to escape a room.

\subsubsection{Reaching for the Light}

First, Bob is in a dark room where the only thing visible is a light bulb, at a 2.5m height, in a small cupboard. Bob hence \textbf{climbs} in the cupboard to touch the bulb, which then turns on the lights. In the physical world, he hence \textbf{goes into \SYSTEM/}, which \textit{remains still} and touches the top of \SYSTEM/'s skeleton.

\subsubsection{The Magic Wall}

Bob then sees a carpet with the words "Start". Once he reaches it, three walls appear. A sign informs him he needs to push them. Bob chooses a wall, but can change his mind and pick another one if he wants. He then has to \textbf{maintain contact} and \textbf{keep pushing} for 10 seconds. The walls' color changes from green to red (accordingly with the timer), to indicate Bob he needs to keep pushing and that a maintained contact is needed. 

When the walls appear, \SYSTEM/ hence \textit{uses the users' intentions-based algorithm} in order to place itself at the chosen wall. When Bob pushes a wall, \SYSTEM/ remains \textit{static}. Once the timer is finished, \SYSTEM/ \textit{steps backwards}, which gives Bob the impression of \textbf{having pushed the wall himself}.


\subsection{Travelling in the Clouds}

After pushing the walls, dust starts flying around Bob, who is then teleported in a forest.

\subsubsection{The Magic Broom}

The user now sees a magic broom. He \textbf{holds} it tightly, and is now \textbf{pulled} by \SYSTEM/, through the forest to the clouds. \SYSTEM/ \textit{pulls the user} with a strong force-feedback, as the broom is actually a cylinder attached to \SYSTEM/ with a string and a spring (see Figure \ref{fig:top_photo} - C).

\subsubsection{Moving in the Clouds}

Once the travel is over, a "Continue" panel appears. When Bob touches it, a chair appears. Bob then \textbf{sits comfortably} in the chair, and \SYSTEM/ \textit{transports} him through the clouds.



\subsubsection{The Ghosts}

The user, in the clouds, is now surrounded by ghosts. He then sees a halo, in which he decides to go into. When he reaches it, he then sees a huge ghost about to go through him. Bob \textbf{remains still} while \SYSTEM/ \textit{initiates the interaction} by \textit{brushing his head with a piece of fabric}.



\section{User study}

The goal of this study is three-fold: (1) validating the implementation of \SYSTEM/, (2) investigating how users experience (i.e. apply and receive) strong forces and (3) collect feedback on the interactions of the demo application.


\textit{Participants.} 8 participants (4 male) aged from 22 to 30 (std = 2.8) volunteered to test the demo application. 4 of the participants were familiar with VR technologies, 2 had only tried VR once and the remaining 2 had never experienced VR.

\textit{Procedure.} Participants were informed they were going to interact with physical props and were asked not to rush within the scene. They were asked to wear an Oculus Rift S HMD as well as Optitrack markers on their dominant hand. They all were introduced to \SYSTEM/ and saw it moving beforehand. A \textit{game master} was present during all the experiments, to ensure the participants' safety and activate some of the interactions. After the experiment, participants filled a Likert-scale questionnaire about their enjoyment on each demo interaction and then participated in a semi-structured interview. They gave approximately 20 minutes of their time.

\subsection{Results}

\subsubsection{Quantitative results.} 
Participants ranked their global enjoyment with a 6.0/7 grade (std = 0.5).

\textit{Favourite interactions.} Users were asked to choose their two favourite interactions in terms of enjoyment, among the five that were provided. 62.5\% of the participants said their favourite interaction was the \textit{transport}, while the remaining 37.5\% preferred the \textit{magic broom} (being pulled). The second favourite interactions were evenly split between \textit{pushing walls, the magic broom, transport and being gone through by ghosts}.

\begin{figure}[h]
    \centering
    \includegraphics[width=\linewidth]{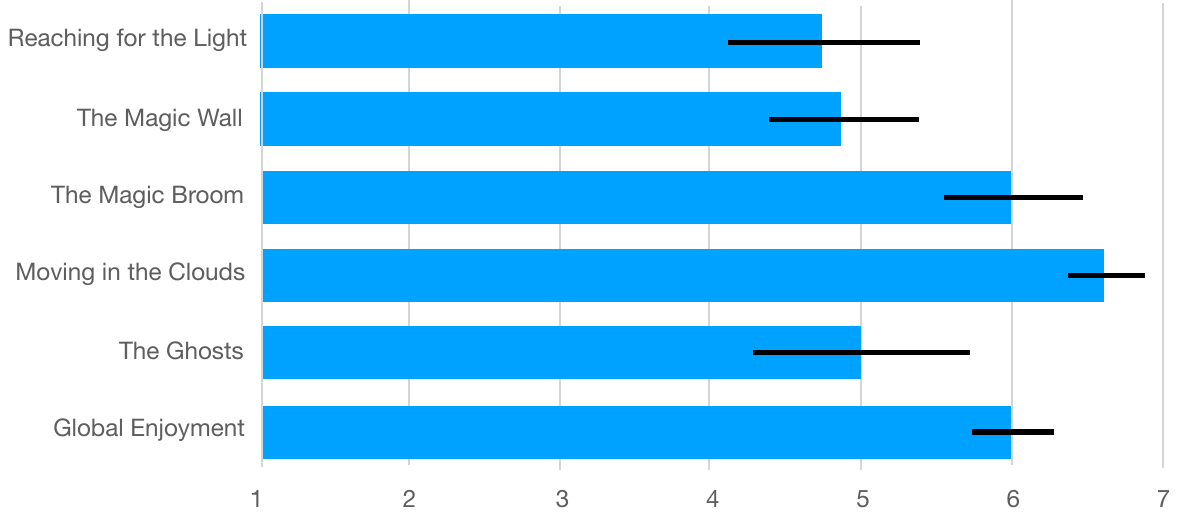}
    \caption{Enjoyment results per interaction, ranked on a 7-point Likert scale - 1 indicates "not enjoyable", 7 indicates "very enjoyable". Error bars indicate the standard deviation of the grades in the users' panel. 
    }
    \label{fig:ResultsUX_Horiz}
\end{figure}

\textit{Force-feedback.} All of the participants ranked the force they applied (\textit{wall}) or applied to them (\textit{broom}) compared to their maximum force on a 7-point Likert scale (1 = pretty soft, 7 = very hard). The forces they applied to the walls was ranked with an average of 5.5 (std = 0.75, min = 5/7, max = 7/7) while the force applied to them with the magic broom was ranked with an average of 6.1 (std=0.64). In particular, 87.5\% of the participants (7/8) ranked the force applied to them with travelling with the magic broom between 6 and 7/7 (the last participant attributed a 5/7 grade).


\textit{Spatial Mismatches.} None of the participants experienced spatial mismatches, even with the non-deterministic scenario involving multiple doors.

\textit{Apprehension.} The participants ranked their fear of being around a moving platform with a Likert scale (1 = not scary at all; 7 = frightening). The average fear was 3.6/7 (std = 1.5). P2 (expert VR user) told us that he would have liked to have noise cancelling ear-puffs and ranked his scare with a 6/7 grade, as the noise was keeping him from being fully immersed. All of the non-expert users ranked their scare with a 2 or 3/7 grade and dove into the VR environment without apprehension.



\subsubsection{Qualitative results.} 

\textit{Whole-body Interactions.} In our semi-structured interview, we discussed the users' game preferences. All of the participants told us that they prefer whole-body interactions in exploration games, where performances do no matter. They all informed us they enjoyed our game and the interactions it provided, and were mostly surprised to be pulled by the broom or transported. 

\textit{Force-feedback.} They were especially surprised by the force provided by the broom, as it was the first \textit{dynamic} interaction they were experiencing. P5 said that she was afraid of heights in the virtual scene, so when the broom started pulling her, she felt quite stressed out. P4 told us she enjoyed the use of passive props and direct manipulation \cite{bryson_direct_2005}. 



\textit{Future Interactions Opportunities.} We asked participants to give us feedback on interactions they would like to experience in VR with \SYSTEM/. Two external expert users told us that they would enjoy climbing on a wall. P4 mentioned virtual escape game, where she could truly benefit from passive haptics, manipulate objects and feel force-feedback. P2 and P6 suggested war games, where they could lean on the walls to get some rest, or hide from enemies. P7 added he would enjoy having more modalities involved, for instance he would appreciate having a sensation of wind when climbing (on a stair or else) to increase his immersion.

\subsection{Discussion}
We now summarize and discuss our main findings.

\textit{CoVR implementation}. The experiment confirmed the robustness of \SYSTEM/ as it did not show any failure during the experiences: Our \textit{robotic system} applied or received strong forces by the participants without damage. Moreover none of the participants experienced collision or spatial mismatches while they were freely walking in the entire room-scale arena thanks to our \textit{trajectory generation} algorithm and more particularly our user-intention model.

\textit{Experiencing strong forces}. The experiment also revealed the benefits of robotic interfaces and more specifically \textit{robotic shape displays} providing strong forces. Indeed, participants spontaneously applied 5.5/7 of their maximum forces when pushing on the walls. One participant reported having applied "very hard" forces (7/7).
Moreover, participants perceived the strong (6.1/7) tension forces when interacting with the broom and enjoyed them (second favourite interaction, and an average of 6/7). Seven participants reported having received "very strong" forces (>6/7).

\textit{Transporting the user}. The favourite interaction was "transport" (6.6/7) where a user was sitting on a chair moving in the VR arena. This interaction requires both a large arena and a robotic system able to displace heavy embedded masses, which are unique features of \SYSTEM/. 

In summary, this experiment revealed that whole-body interactions involving strong forces (applying forces, receiving forces or embedding heavy masses) are a promising direction for future \textit{Robotic Shape Displays}. 





\section{Conclusion and Future Work}

We presented \SYSTEM/ a novel \textit{Robotic Shape Display} for room-scale VR arena providing whole-body interactions and strong force feedback. We also proposed a low-computational user intention model compatible with common HMDs to support non-deterministic scenarios. The technical evaluation and the user study demonstrated the feasability of the approach, its usability and the relevance of interactions involving strong forces. While \SYSTEM/ addresses several interactions and technical challenges, we see several directions for future work. 

\textit{Adding multiple columns}. A main limitation of our current setup is the use of a single column in the VR arena. One approach consists of mounting additional 2D Cartesian robots on the sides of the VR arena to control horizontal columns. Another one is to add a second ceiling robot (the robots can share the rails). These two approaches limit the work-area of the additional columns but appropriate control strategies can optimize trajectories and augment interaction possibilities, especially with multiple users (see below). 

\textit{Combining multiple RSDs}. Our approach is compatible with previous Robotic Shape Display solutions. For instance, we envision a VR arena combining \SYSTEM/ with a swarm of mobile robots such as \cite{suzuki_roomshift_2020, he_physhare:_2017, wang_movevr_2020}. These ones can collect objects on the ground and bring them back to \SYSTEM/. Our \SYSTEM/'s trajectory generation algorithm remains valid in such configurations. More DoFs could also be integrated to \SYSTEM/ by coupling it with a Kuka robotic arm or a Snake Charmer \cite{araujo_snake_2016} interface.




\textit{Augmenting I/O capabilities}. We will investigate how additional capabilities can improve user experience. For instance, it would be interesting to augment a column with sensors (e.g. touch input, force sensors, proximity sensors, etc.).Adding a depth camera could enable the detection of untracked moving bodies, such as an unexpected pet in the VR arena.
Haptic stimuli can be expanded to vibrations, sliding, textures, temperatures, or to shape changing illusions. For instance, heat-lamps or wind-blowers could also be integrated \cite{shaw_heat_2019}.

\textit{Collaboration}. Finally, our system is currently designed for a single user interaction. We plan to investigate remote-presence interaction: a second identical structure can for instance be assembled in another room. Users in each room can interact with different VOIs, share mutual physical contact or collaboratively manipulate objects \cite{brave_tangible_1998, he_physhare:_2017}. We also plan to investigate which scenarios (e.g. a master and a slave) and which interactions would support collaborative interaction in a single arena. Our software implementation can already support several users in the same arena - each user being considered as an obstacle, but collaborative interactions raise multiple challenges \cite{he_collabovr:_2019}.

\section{Acknowledgements}
We would like to thank M. Teyssier, C. Rigaud, B. Geslain, S. Sakr, M. Serrano, J. Müller, J. Gugenheimer as well as the participants of the studies. 
\balance

\bibliographystyle{SIGCHI-Reference-Format}
\bibliography{references}


\begin{thebibliography}{00}


\ifx \showCODEN    \undefined \def \showCODEN     #1{\unskip}     \fi
\ifx \showDOI      \undefined \def \showDOI       #1{{\tt DOI:}\penalty0{#1}\ }
  \fi
\ifx \showISBNx    \undefined \def \showISBNx     #1{\unskip}     \fi
\ifx \showISBNxiii \undefined \def \showISBNxiii  #1{\unskip}     \fi
\ifx \showISSN     \undefined \def \showISSN      #1{\unskip}     \fi
\ifx \showLCCN     \undefined \def \showLCCN      #1{\unskip}     \fi
\ifx \shownote     \undefined \def \shownote      #1{#1}          \fi
\ifx \showarticletitle \undefined \def \showarticletitle #1{#1}   \fi
\ifx \showURL      \undefined \def \showURL       #1{#1}          \fi

\bibitem{noauthor_cybergrasp_2019}
 2019a.
\newblock {CyberGrasp}.
\newblock   (2019).
\newblock
\showURL{%
\url{http://www.cyberglovesystems.com/cybergrasp}}


\bibitem{noauthor_oculus_2019}
 2019b.
\newblock Oculus {Rift} {S}.
\newblock   (2019).
\newblock
\showURL{%
\url{https://www.oculus.com/rift-s/?locale=fr_FR}}


\bibitem{noauthor_steamvr_2019}
 2019c.
\newblock {SteamVR} - {Valve} {Corporation}.
\newblock   (2019).
\newblock
\showURL{%
\url{https://www.steamvr.com/en/}}


\bibitem{noauthor_teslasuit_2019}
 2019d.
\newblock Teslasuit {\textbar} {Full} body haptic {VR} suit for motion capture
  and training.
\newblock   (2019).
\newblock
\showURL{%
\url{https://teslasuit.io/}}


\bibitem{abtahi_beyond_2019}
{Parastoo Abtahi}, {Benoit Landry}, {Jackie~(Junrui) Yang}, {Marco Pavone},
  {Sean Follmer}, {and} {James~A. Landay}. 2019.
\newblock \showarticletitle{Beyond {The} {Force}: {Using} {Quadcopters} to
  {Appropriate} {Objects} and the {Environment} for {Haptics} in {Virtual}
  {Reality}}. In {\em Proceedings of the 2019 {CHI} {Conference} on {Human}
  {Factors} in {Computing} {Systems} - {CHI} '19}. ACM Press, Glasgow, Scotland
  Uk, 1--13.
\newblock
\showISBNx{978-1-4503-5970-2}
\showDOI{%
\url{http://dx.doi.org/10.1145/3290605.3300589}}


\bibitem{amirpour_design_2019}
{E. Amirpour}, {M. Savabi}, {A. Saboukhi}, {M.~Rahimi Gorii}, {H. Ghafarirad},
  {R. Fesharakifard}, {and} {S.~Mehdi Rezaei}. 2019.
\newblock \showarticletitle{Design and {Optimization} of a {Multi}-{DOF} {Hand}
  {Exoskeleton} for {Haptic} {Applications}}. In {\em 2019 7th {International}
  {Conference} on {Robotics} and {Mechatronics} ({ICRoM})}. 270--275.
\newblock
\showDOI{%
\url{http://dx.doi.org/10.1109/ICRoM48714.2019.9071884}}
\newblock
\shownote{ISSN: 2572-6889.}


\bibitem{araujo_snake_2016}
{Bruno Araujo}, {Ricardo Jota}, {Varun Perumal}, {Jia~Xian Yao}, {Karan Singh},
  {and} {Daniel Wigdor}. 2016.
\newblock \showarticletitle{Snake {Charmer}: {Physically} {Enabling} {Virtual}
  {Objects}}. In {\em Proceedings of the {TEI} '16: {Tenth} {International}
  {Conference} on {Tangible}, {Embedded}, and {Embodied} {Interaction} - {TEI}
  '16}. ACM Press, Eindhoven, Netherlands, 218--226.
\newblock
\showISBNx{978-1-4503-3582-9}
\showDOI{%
\url{http://dx.doi.org/10.1145/2839462.2839484}}


\bibitem{auda_around_2019}
{Jonas Auda}, {Max Pascher}, {and} {Stefan Schneegass}. 2019.
\newblock \showarticletitle{Around the ({Virtual}) {World}: {Infinite}
  {Walking} in {Virtual} {Reality} {Using} {Electrical} {Muscle}
  {Stimulation}}. In {\em Proceedings of the 2019 {CHI} {Conference} on {Human}
  {Factors} in {Computing} {Systems} - {CHI} '19}. ACM Press, Glasgow, Scotland
  Uk, 1--8.
\newblock
\showISBNx{978-1-4503-5970-2}
\showDOI{%
\url{http://dx.doi.org/10.1145/3290605.3300661}}


\bibitem{azmandian_haptic_2016-1}
{Mahdi Azmandian}, {Mark Hancock}, {Hrvoje Benko}, {Eyal Ofek}, {and}
  {Andrew~D. Wilson}. 2016a.
\newblock \showarticletitle{Haptic {Retargeting}: {Dynamic} {Repurposing} of
  {Passive} {Haptics} for {Enhanced} {Virtual} {Reality} {Experiences}}. In
  {\em Proceedings of the 2016 {CHI} {Conference} on {Human} {Factors} in
  {Computing} {Systems} - {CHI} '16}. ACM Press, Santa Clara, California, USA,
  1968--1979.
\newblock
\showISBNx{978-1-4503-3362-7}
\showDOI{%
\url{http://dx.doi.org/10.1145/2858036.2858226}}


\bibitem{azmandian_haptic_2016}
{Mahdi Azmandian}, {Mark Hancock}, {Hrvoje Benko}, {Eyal Ofek}, {and}
  {Andrew~D. Wilson}. 2016b.
\newblock \showarticletitle{Haptic {Retargeting}: {Dynamic} {Repurposing} of
  {Passive} {Haptics} for {Enhanced} {Virtual} {Reality} {Experiences}}. In
  {\em Proceedings of the 2016 {CHI} {Conference} on {Human} {Factors} in
  {Computing} {Systems} - {CHI} '16}. ACM Press, Santa Clara, California, USA,
  1968--1979.
\newblock
\showISBNx{978-1-4503-3362-7}
\showDOI{%
\url{http://dx.doi.org/10.1145/2858036.2858226}}


\bibitem{benko_normaltouch_2016}
{Hrvoje Benko}, {Christian Holz}, {Mike Sinclair}, {and} {Eyal Ofek}. 2016.
\newblock \showarticletitle{{NormalTouch} and {TextureTouch}: {High}-fidelity
  {3D} {Haptic} {Shape} {Rendering} on {Handheld} {Virtual} {Reality}
  {Controllers}}. In {\em Proceedings of the 29th {Annual} {Symposium} on
  {User} {Interface} {Software} and {Technology} - {UIST} '16}. ACM Press,
  Tokyo, Japan, 717--728.
\newblock
\showISBNx{978-1-4503-4189-9}
\showDOI{%
\url{http://dx.doi.org/10.1145/2984511.2984526}}


\bibitem{binsted_eyehand_2001}
{Gordon Binsted}, {Romeo Chua}, {Werner Helsen}, {and} {Digby Elliott}. 2001.
\newblock \showarticletitle{Eye–hand coordination in goal-directed aiming}.
\newblock {\em Human Movement Science\/} {20}, 4-5 (Nov. 2001), 563--585.
\newblock
\showISSN{01679457}
\showDOI{%
\url{http://dx.doi.org/10.1016/S0167-9457(01)00068-9}}


\bibitem{brave_tangible_1998}
{Scott Brave}, {Hiroshi Ishii}, {and} {Andrew Dahley}. 1998.
\newblock \showarticletitle{Tangible interfaces for remote collaboration and
  communication}. In {\em Proceedings of the 1998 {ACM} conference on
  {Computer} supported cooperative work - {CSCW} '98}. ACM Press, Seattle,
  Washington, United States, 169--178.
\newblock
\showISBNx{978-1-58113-009-6}
\showDOI{%
\url{http://dx.doi.org/10.1145/289444.289491}}


\bibitem{bryson_direct_2005}
{Steve Bryson}. 2005.
\newblock \showarticletitle{Direct {Manipulation} in {Virtual} {Reality}}.
\newblock In {\em Visualization {Handbook}}. Elsevier, 413--430.
\newblock
\showISBNx{978-0-12-387582-2}
\showDOI{%
\url{http://dx.doi.org/10.1016/B978-012387582-2/50023-X}}


\bibitem{cheng_iturk:_2018}
{Lung-Pan Cheng}, {Li Chang}, {Sebastian Marwecki}, {and} {Patrick Baudisch}.
  2018.
\newblock \showarticletitle{{iTurk}: {Turning} {Passive} {Haptics} into
  {Active} {Haptics} by {Making} {Users} {Reconfigure} {Props} in {Virtual}
  {Reality}}. In {\em Proceedings of the 2018 {CHI} {Conference} on {Human}
  {Factors} in {Computing} {Systems} - {CHI} '18}. ACM Press, Montreal QC,
  Canada, 1--10.
\newblock
\showISBNx{978-1-4503-5620-6}
\showDOI{%
\url{http://dx.doi.org/10.1145/3173574.3173663}}


\bibitem{cheng_haptic_2014}
{Lung-Pan Cheng}, {Patrick Lühne}, {Pedro Lopes}, {Christoph Sterz}, {and}
  {Patrick Baudisch}. 2014.
\newblock \showarticletitle{Haptic {Turk}: a {Motion} {Platform} {Based} on
  {People}}.
\newblock  (2014), 11.
\newblock


\bibitem{cheng_mutual_2017}
{Lung-Pan Cheng}, {Sebastian Marwecki}, {and} {Patrick Baudisch}. 2017a.
\newblock \showarticletitle{Mutual {Human} {Actuation}}. In {\em Proceedings of
  the 30th {Annual} {ACM} {Symposium} on {User} {Interface} {Software} and
  {Technology} - {UIST} '17}. ACM Press, Qu\&\#233;bec City, QC, Canada,
  797--805.
\newblock
\showISBNx{978-1-4503-4981-9}
\showDOI{%
\url{http://dx.doi.org/10.1145/3126594.3126667}}


\bibitem{cheng_sparse_2017}
{Lung-Pan Cheng}, {Eyal Ofek}, {Christian Holz}, {Hrvoje Benko}, {and}
  {Andrew~D. Wilson}. 2017b.
\newblock \showarticletitle{Sparse {Haptic} {Proxy}: {Touch} {Feedback} in
  {Virtual} {Environments} {Using} a {General} {Passive} {Prop}}. In {\em
  Proceedings of the 2017 {CHI} {Conference} on {Human} {Factors} in
  {Computing} {Systems} - {CHI} '17}. ACM Press, Denver, Colorado, USA,
  3718--3728.
\newblock
\showISBNx{978-1-4503-4655-9}
\showDOI{%
\url{http://dx.doi.org/10.1145/3025453.3025753}}


\bibitem{cheng_turkdeck:_2015}
{Lung-Pan Cheng}, {Thijs Roumen}, {Hannes Rantzsch}, {Sven Köhler}, {Patrick
  Schmidt}, {Robert Kovacs}, {Johannes Jasper}, {Jonas Kemper}, {and} {Patrick
  Baudisch}. 2015.
\newblock \showarticletitle{{TurkDeck}: {Physical} {Virtual} {Reality} {Based}
  on {People}}. In {\em Proceedings of the 28th {Annual} {ACM} {Symposium} on
  {User} {Interface} {Software} \& {Technology} - {UIST} '15}. ACM Press,
  Daegu, Kyungpook, Republic of Korea, 417--426.
\newblock
\showISBNx{978-1-4503-3779-3}
\showDOI{%
\url{http://dx.doi.org/10.1145/2807442.2807463}}


\bibitem{choi_grabity_2017}
{Inrak Choi}, {Heather Culbertson}, {Mark~R. Miller}, {Alex Olwal}, {and} {Sean
  Follmer}. 2017.
\newblock \showarticletitle{Grabity: {A} {Wearable} {Haptic} {Interface} for
  {Simulating} {Weight} and {Grasping} in {Virtual} {Reality}}. In {\em
  Proceedings of the 30th {Annual} {ACM} {Symposium} on {User} {Interface}
  {Software} and {Technology} - {UIST} '17}. ACM Press, Qu\&\#233;bec City, QC,
  Canada, 119--130.
\newblock
\showISBNx{978-1-4503-4981-9}
\showDOI{%
\url{http://dx.doi.org/10.1145/3126594.3126599}}


\bibitem{choi_claw:_2018}
{Inrak Choi}, {Eyal Ofek}, {Hrvoje Benko}, {Mike Sinclair}, {and} {Christian
  Holz}. 2018.
\newblock \showarticletitle{{CLAW}: {A} {Multifunctional} {Handheld} {Haptic}
  {Controller} for {Grasping}, {Touching}, and {Triggering} in {Virtual}
  {Reality}}. In {\em Proceedings of the 2018 {CHI} {Conference} on {Human}
  {Factors} in {Computing} {Systems} - {CHI} '18}. ACM Press, Montreal QC,
  Canada, 1--13.
\newblock
\showISBNx{978-1-4503-5620-6}
\showDOI{%
\url{http://dx.doi.org/10.1145/3173574.3174228}}


\bibitem{de_tinguy_weatavix_2020}
{Xavier de Tinguy}, {Thomas Howard}, {Claudio Pacchierotti}, {Maud Marchal},
  {and} {Anatole Lécuyer}. 2020.
\newblock \showarticletitle{{WeATaViX}: {WEarable} {Actuated} {TAngibles} for
  {VIrtual} reality {eXperiences}}.
\newblock  (2020), 9.
\newblock


\bibitem{force_dimension_force_2019}
{Force Dimension}. 2019.
\newblock Force {Dimension} - products.
\newblock   (2019).
\newblock
\showURL{%
\url{http://www.forcedimension.com/}}


\bibitem{formaglio_performance_2005}
{A. Formaglio}, {A. Giannitrapani}, {M. Franzini}, {D. Prattichizzo}, {and} {F.
  Barbagli}. 2005.
\newblock \showarticletitle{Performance of {Mobile} {Haptic} {Interfaces}}. In
  {\em Proceedings of the 44th {IEEE} {Conference} on {Decision} and
  {Control}}. 8343--8348.
\newblock
\showDOI{%
\url{http://dx.doi.org/10.1109/CDC.2005.1583513}}


\bibitem{gosselin_widening_2007}
{F. Gosselin}, {C. Andriot}, {F. Bergez}, {and} {X. Merlhiot}. 2007.
\newblock \showarticletitle{Widening 6-{DOF} haptic devices workspace with an
  additional degree of freedom}. In {\em Second {Joint} {EuroHaptics}
  {Conference} and {Symposium} on {Haptic} {Interfaces} for {Virtual}
  {Environment} and {Teleoperator} {Systems} ({WHC}'07)}. 452--457.
\newblock
\showDOI{%
\url{http://dx.doi.org/10.1109/WHC.2007.127}}


\bibitem{haption_scale1_2019}
{Haption}. 2019a.
\newblock Scale1™ - {HAPTION} {SA}.
\newblock   (2019).
\newblock
\showURL{%
\url{https://www.haption.com/en/products-en/scale-one-en.html}}


\bibitem{haption_virtuose_2019}
{Haption}. 2019b.
\newblock Virtuose™ {6D} - {HAPTION} {SA}.
\newblock   (2019).
\newblock
\showURL{%
\url{https://www.haption.com/en/products-en/virtuose-6d-en.html}}


\bibitem{he_collabovr:_2019}
{Zhenyi He} {and} {Ken Perlin}. 2019.
\newblock \showarticletitle{{CollaboVR}: {A} {Reconfigurable} {Framework} for
  {Multi}-user to {Communicate} in {Virtual} {Reality}}.
\newblock {\em arXiv:1912.03863 [cs]\/} (Dec. 2019).
\newblock
\showURL{%
\url{http://arxiv.org/abs/1912.03863}}
\newblock
\shownote{arXiv: 1912.03863.}


\bibitem{he_robotic_2017}
{Zhenyi He}, {Fengyuan Zhu}, {Aaron Gaudette}, {and} {Ken Perlin}. 2017b.
\newblock \showarticletitle{Robotic {Haptic} {Proxies} for {Collaborative}
  {Virtual} {Reality}}.
\newblock {\em arXiv:1701.08879 [cs]\/} (Jan. 2017).
\newblock
\showURL{%
\url{http://arxiv.org/abs/1701.08879}}
\newblock
\shownote{arXiv: 1701.08879.}


\bibitem{he_physhare:_2017}
{Zhenyi He}, {Fengyuan Zhu}, {and} {Ken Perlin}. 2017a.
\newblock \showarticletitle{{PhyShare}: {Sharing} {Physical} {Interaction} in
  {Virtual} {Reality}}.
\newblock {\em arXiv:1708.04139 [cs]\/} (Aug. 2017).
\newblock
\showURL{%
\url{http://arxiv.org/abs/1708.04139}}
\newblock
\shownote{arXiv: 1708.04139.}


\bibitem{heo_thors_2018}
{Seongkook Heo}, {Christina Chung}, {Geehyuk Lee}, {and} {Daniel Wigdor}. 2018.
\newblock \showarticletitle{Thor's {Hammer}: {An} {Ungrounded} {Force}
  {Feedback} {Device} {Utilizing} {Propeller}-{Induced} {Propulsive} {Force}}.
  In {\em Proceedings of the 2018 {CHI} {Conference} on {Human} {Factors} in
  {Computing} {Systems} - {CHI} '18}. ACM Press, Montreal QC, Canada, 1--11.
\newblock
\showISBNx{978-1-4503-5620-6}
\showDOI{%
\url{http://dx.doi.org/10.1145/3173574.3174099}}


\bibitem{hettiarachchi_annexing_2016}
{Anuruddha Hettiarachchi} {and} {Daniel Wigdor}. 2016.
\newblock \showarticletitle{Annexing {Reality}: {Enabling} {Opportunistic}
  {Use} of {Everyday} {Objects} as {Tangible} {Proxies} in {Augmented}
  {Reality}}. In {\em Proceedings of the 2016 {CHI} {Conference} on {Human}
  {Factors} in {Computing} {Systems} - {CHI} '16}. ACM Press, Santa Clara,
  California, USA, 1957--1967.
\newblock
\showISBNx{978-1-4503-3362-7}
\showDOI{%
\url{http://dx.doi.org/10.1145/2858036.2858134}}


\bibitem{hoppe_vrhapticdrones:_2018}
{Matthias Hoppe}, {Pascal Knierim}, {Thomas Kosch}, {Markus Funk}, {Lauren
  Futami}, {Stefan Schneegass}, {Niels Henze}, {Albrecht Schmidt}, {and} {Tonja
  Machulla}. 2018.
\newblock \showarticletitle{{VRHapticDrones}: {Providing} {Haptics} in
  {Virtual} {Reality} through {Quadcopters}}. In {\em Proceedings of the 17th
  {International} {Conference} on {Mobile} and {Ubiquitous} {Multimedia} -
  {MUM} 2018}. ACM Press, Cairo, Egypt, 7--18.
\newblock
\showISBNx{978-1-4503-6594-9}
\showDOI{%
\url{http://dx.doi.org/10.1145/3282894.3282898}}


\bibitem{hoppe_human_2020}
{Matthias Hoppe}, {Beat Rossmy}, {Daniel~Peter Neumann}, {Stephan Streuber},
  {Albrecht Schmidt}, {and} {Tonja-Katrin Machulla}. 2020.
\newblock \showarticletitle{A {Human} {Touch}: {Social} {Touch} {Increases} the
  {Perceived} {Human}-likeness of {Agents} in {Virtual} {Reality}}.
\newblock  (2020), 11.
\newblock


\bibitem{hoshino_contruction_1995}
{Hiroshi Hoshino}. 1995.
\newblock A {Contruction} {MEthod} of {Virtual} {Haptic} {Space}.
\newblock   (1995).
\newblock
\showURL{%
\url{http://www.files.tachilab.org/publications/intconf1900/hoshino199511ICAT.pdf}}


\bibitem{hsin-yu_huang_haptic-go-round_2020}
{Hsin-Yu Huang}, {Chih-Wei Ning}, {Po-Yao~(Cosmos) Wang}, {Jen-Hao Cheng},
  {and} {Lung-Pan Cheng}. 2020.
\newblock Haptic-go-round: {A} surrounding {Platform} for {EncounterType}
  {Haptic} in {VR} experienes.
\newblock   (2020).
\newblock
\showURL{%
\url{https://dl.acm.org/doi/pdf/10.1145/3334480.3383136}}


\bibitem{insko_passive_2001}
{Brent~Edward Insko}. 2001.
\newblock \showarticletitle{Passive {Haptics} {Significantly} {Enhances}
  {Virtual} {Environments}}.
\newblock  (2001), 111.
\newblock


\bibitem{kim_encountered-type_2018}
{Yaesol Kim}, {Hyun~Jung Kim}, {and} {Young~J. Kim}. 2018.
\newblock \showarticletitle{Encountered-type haptic display for large {VR}
  environment using per-plane reachability maps: {Encountered}-type {Haptic}
  {Display} for {Large} {VR} {Environment}}.
\newblock {\em Computer Animation and Virtual Worlds\/} {29}, 3-4 (May 2018),
  e1814.
\newblock
\showISSN{15464261}
\showDOI{%
\url{http://dx.doi.org/10.1002/cav.1814}}


\bibitem{knierim_tactile_2017}
{Pascal Knierim}, {Thomas Kosch}, {Valentin Schwind}, {Markus Funk}, {Francisco
  Kiss}, {Stefan Schneegass}, {and} {Niels Henze}. 2017.
\newblock \showarticletitle{Tactile {Drones} - {Providing} {Immersive}
  {Tactile} {Feedback} in {Virtual} {Reality} through {Quadcopters}}. In {\em
  Proceedings of the 2017 {CHI} {Conference} {Extended} {Abstracts} on {Human}
  {Factors} in {Computing} {Systems} - {CHI} {EA} '17}. ACM Press, Denver,
  Colorado, USA, 433--436.
\newblock
\showISBNx{978-1-4503-4656-6}
\showDOI{%
\url{http://dx.doi.org/10.1145/3027063.3050426}}


\bibitem{lee_mobile_2007}
{Chaehyun Lee}, {Min~Sik Hong}, {In Lee}, {Oh~Kyu Choi}, {Kyung-Lyong Han},
  {Yoo~Yeon Kim}, {Seungmoon Choi}, {and} {Jin~S Lee}. 2007.
\newblock \showarticletitle{Mobile {Haptic} {Interface} for {Large} {Immersive}
  {Virtual} {Environments}: {PoMHI} v0.5}.
\newblock  (2007), 2.
\newblock


\bibitem{lee_system_2009}
{In Lee}, {Inwook Hwang}, {Kyung-Lyoung Han}, {Oh~Kyu Choi}, {Seungmoon Choi},
  {and} {Jin~S. Lee}. 2009.
\newblock \showarticletitle{System improvements in {Mobile} {Haptic}
  {Interface}}. In {\em World {Haptics} 2009 - {Third} {Joint} {EuroHaptics}
  conference and {Symposium} on {Haptic} {Interfaces} for {Virtual}
  {Environment} and {Teleoperator} {Systems}}. IEEE, Salt Lake City, UT, USA,
  109--114.
\newblock
\showISBNx{978-1-4244-3858-7}
\showDOI{%
\url{http://dx.doi.org/10.1109/WHC.2009.4810834}}


\bibitem{lee_torc:_2019}
{Jaeyeon Lee}, {Mike Sinclair}, {Mar Gonzalez-Franco}, {Eyal Ofek}, {and}
  {Christian Holz}. 2019.
\newblock \showarticletitle{{TORC}: {A} {Virtual} {Reality} {Controller} for
  {In}-{Hand} {High}-{Dexterity} {Finger} {Interaction}}. In {\em Proceedings
  of the 2019 {CHI} {Conference} on {Human} {Factors} in {Computing} {Systems}
  - {CHI} '19}. ACM Press, Glasgow, Scotland Uk, 1--13.
\newblock
\showISBNx{978-1-4503-5970-2}
\showDOI{%
\url{http://dx.doi.org/10.1145/3290605.3300301}}


\bibitem{lepecq_afforded_2008}
{Jean-Claude Lepecq}, {Lionel Bringoux}, {Jean-Marie Pergandi}, {Thelma Coyle},
  {and} {Daniel Mestre}. 2008.
\newblock \showarticletitle{Afforded {Actions} as a {Behavioral} {Assessment}
  of {Physical} {Presence}}.
\newblock  (2008), 8.
\newblock


\bibitem{lopes_impacto:_2015}
{Pedro Lopes}, {Alexandra Ion}, {and} {Patrick Baudisch}. 2015.
\newblock \showarticletitle{Impacto: {Simulating} {Physical} {Impact} by
  {Combining} {Tactile} {Stimulation} with {Electrical} {Muscle}
  {Stimulation}}. In {\em Proceedings of the 28th {Annual} {ACM} {Symposium} on
  {User} {Interface} {Software} \& {Technology} - {UIST} '15}. ACM Press,
  Daegu, Kyungpook, Republic of Korea, 11--19.
\newblock
\showISBNx{978-1-4503-3779-3}
\showDOI{%
\url{http://dx.doi.org/10.1145/2807442.2807443}}


\bibitem{mcneely_robotic_1993}
{W.~A. McNeely}. 1993.
\newblock \showarticletitle{Robotic graphics: a new approach to force feedback
  for virtual reality}. In {\em Proceedings of {IEEE} {Virtual} {Reality}
  {Annual} {International} {Symposium}}. 336--341.
\newblock
\showDOI{%
\url{http://dx.doi.org/10.1109/VRAIS.1993.380761}}


\bibitem{mercado_design_2020}
{Víctor Mercado} {and} {Univ Rennes}. 2020.
\newblock \showarticletitle{Design and {Evaluation} of {Interaction}
  {Techniques} {Dedicated} to {Integrate} {Encountered}-{Type} {Haptic}
  {Displays} in {Virtual} {Environments}}.
\newblock  (2020), 9.
\newblock


\bibitem{minamizawa_simplified_2010}
{Kouta Minamizawa}, {Domenico Prattichizzo}, {and} {Susumu Tachi}. 2010.
\newblock \showarticletitle{Simplified design of haptic display by extending
  one-point kinesthetic feedback to multipoint tactile feedback}. In {\em 2010
  {IEEE} {Haptics} {Symposium}}. IEEE, Waltham, MA, USA, 257--260.
\newblock
\showISBNx{978-1-4244-6821-8}
\showDOI{%
\url{http://dx.doi.org/10.1109/HAPTIC.2010.5444646}}


\bibitem{nitzsche_design_2003}
{Norbert Nitzsche}, {Uwe~D. Hanebeck}, {and} {G. Schmidt}. 2003.
\newblock \showarticletitle{Design issues of mobile haptic interfaces}.
\newblock {\em Journal of Robotic Systems\/} {20}, 9 (Sept. 2003), 549--556.
\newblock
\showISSN{0741-2223, 1097-4563}
\showDOI{%
\url{http://dx.doi.org/10.1002/rob.10105}}


\bibitem{pair_flatworld:_2003}
{J. Pair}, {U. Neumann}, {D. Piepol}, {and} {B. Swartout}. 2003.
\newblock \showarticletitle{{FlatWorld}: combining {Hollywood} set-design
  techniques with {VR}}.
\newblock {\em IEEE Computer Graphics and Applications\/} {23}, 1 (Jan. 2003),
  12--15.
\newblock
\showISSN{0272-1716}
\showDOI{%
\url{http://dx.doi.org/10.1109/MCG.2003.1159607}}


\bibitem{razzaque_eurographics_2001}
{Sharif Razzaque}, {Zachariah Kohn}, {and} {Mary~C. Whitton}. 2001.
\newblock {\em {EUROGRAPHICS} 2001 / {Jonathan} {C}. {Roberts} {Short}
  {Presentation} © {The} {Eurographics} {Association} 2001. {Redirected}
  {Walking}}.
\newblock


\bibitem{rogers_exploring_2019}
{Katja Rogers}, {Jana Funke}, {Julian Frommel}, {Sven Stamm}, {and} {Michael
  Weber}. 2019.
\newblock \showarticletitle{Exploring {Interaction} {Fidelity} in {Virtual}
  {Reality}: {Object} {Manipulation} and {Whole}-{Body} {Movements}}. In {\em
  Proceedings of the 2019 {CHI} {Conference} on {Human} {Factors} in
  {Computing} {Systems} - {CHI} '19}. ACM Press, Glasgow, Scotland Uk, 1--14.
\newblock
\showISBNx{978-1-4503-5970-2}
\showDOI{%
\url{http://dx.doi.org/10.1145/3290605.3300644}}


\bibitem{satler_control_2011}
{Massimo Satler}, {Carlo~A. Avizzano}, {and} {Emanuele Ruffaldi}. 2011.
\newblock \showarticletitle{Control of a desktop mobile haptic interface}. In
  {\em 2011 {IEEE} {World} {Haptics} {Conference}}. IEEE, Istanbul, 415--420.
\newblock
\showISBNx{978-1-4577-0299-0}
\showDOI{%
\url{http://dx.doi.org/10.1109/WHC.2011.5945522}}


\bibitem{seifi_haptipedia:_2019}
{Hasti Seifi}, {Farimah Fazlollahi}, {Michael Oppermann}, {John~Andrew
  Sastrillo}, {Jessica Ip}, {Ashutosh Agrawal}, {Gunhyuk Park}, {Katherine~J.
  Kuchenbecker}, {and} {Karon~E. MacLean}. 2019.
\newblock \showarticletitle{Haptipedia: {Accelerating} {Haptic} {Device}
  {Discovery} to {Support} {Interaction} \& {Engineering} {Design}}. In {\em
  Proceedings of the 2019 {CHI} {Conference} on {Human} {Factors} in
  {Computing} {Systems} - {CHI} '19}. ACM Press, Glasgow, Scotland Uk, 1--12.
\newblock
\showISBNx{978-1-4503-5970-2}
\showDOI{%
\url{http://dx.doi.org/10.1145/3290605.3300788}}


\bibitem{shaw_heat_2019}
{Emily Shaw}, {Tessa Roper}, {Tommy Nilsson}, {Glyn Lawson}, {Sue V.~G. Cobb},
  {and} {Daniel Miller}. 2019.
\newblock \showarticletitle{The {Heat} is {On}: {Exploring} {User} {Behaviour}
  in a {Multisensory} {Virtual} {Environment} for {Fire} {Evacuation}}.
\newblock {\em Proceedings of the 2019 CHI Conference on Human Factors in
  Computing Systems - CHI '19\/} (2019), 1--13.
\newblock
\showDOI{%
\url{http://dx.doi.org/10.1145/3290605.3300856}}
\newblock
\shownote{arXiv: 1902.04573.}


\bibitem{shigeta_motion_2007}
{Ken Shigeta}, {Yuji Sato}, {and} {Yasuyoshi Yokokohji}. 2007.
\newblock \showarticletitle{Motion {Planning} of {Encountered}-type {Haptic}
  {Device} for {Multiple} {Fingertips} {Based} on {Minimum} {Distance} {Point}
  {Information}}. In {\em Second {Joint} {EuroHaptics} {Conference} and
  {Symposium} on {Haptic} {Interfaces} for {Virtual} {Environment} and
  {Teleoperator} {Systems} ({WHC}'07)}. IEEE, Tsukaba, 188--193.
\newblock
\showISBNx{978-0-7695-2738-3}
\showDOI{%
\url{http://dx.doi.org/10.1109/WHC.2007.85}}


\bibitem{siu_shapeshift:_2018}
{Alexa~F. Siu}, {Eric~J. Gonzalez}, {Shenli Yuan}, {Jason~B. Ginsberg}, {and}
  {Sean Follmer}. 2018.
\newblock \showarticletitle{{shapeShift}: {2D} {Spatial} {Manipulation} and
  {Self}-{Actuation} of {Tabletop} {Shape} {Displays} for {Tangible} and
  {Haptic} {Interaction}}. In {\em Proceedings of the 2018 {CHI} {Conference}
  on {Human} {Factors} in {Computing} {Systems} - {CHI} '18}. ACM Press,
  Montreal QC, Canada, 1--13.
\newblock
\showISBNx{978-1-4503-5620-6}
\showDOI{%
\url{http://dx.doi.org/10.1145/3173574.3173865}}


\bibitem{strasnick_haptic_2018}
{Evan Strasnick}, {Christian Holz}, {Eyal Ofek}, {Mike Sinclair}, {and} {Hrvoje
  Benko}. 2018.
\newblock \showarticletitle{Haptic {Links}: {Bimanual} {Haptics} for {Virtual}
  {Reality} {Using} {Variable} {Stiffness} {Actuation}}. In {\em Proceedings of
  the 2018 {CHI} {Conference} on {Human} {Factors} in {Computing} {Systems} -
  {CHI} '18}. ACM Press, Montreal QC, Canada, 1--12.
\newblock
\showISBNx{978-1-4503-5620-6}
\showDOI{%
\url{http://dx.doi.org/10.1145/3173574.3174218}}


\bibitem{suzuki_roomshift_2020}
{Ryo Suzuki}, {Hooman Hedayati}, {Clement Zheng}, {James Bohn}, {Daniel
  Szafir}, {Ellen Yi-Luen Do}, {Mark~D Gross}, {and} {Daniel Leithinger}.
  2020a.
\newblock \showarticletitle{{RoomShift}: {Room}-scale {Dynamic} {Haptics} for
  {VR} with {Furniture}-moving {Swarm} {Robots}}.
\newblock  (2020), 11.
\newblock


\bibitem{suzuki_lifttiles_2020}
{Ryo Suzuki}, {Ryosuke Nakayama}, {Dan Liu}, {Yasuaki Kakehi}, {Mark~D. Gross},
  {and} {Daniel Leithinger}. 2020b.
\newblock \showarticletitle{{LiftTiles}: {Constructive} {Building} {Blocks} for
  {Prototyping} {Room}-scale {Shape}-changing {Interfaces}}. In {\em
  Proceedings of the {Fourteenth} {International} {Conference} on {Tangible},
  {Embedded}, and {Embodied} {Interaction}}. ACM, Sydney NSW Australia,
  143--151.
\newblock
\showISBNx{978-1-4503-6107-1}
\showDOI{%
\url{http://dx.doi.org/10.1145/3374920.3374941}}


\bibitem{teng_aarnio_2019}
{Shan-Yuan Teng}, {Da-Yuan Huang}, {Chi Wang}, {Jun Gong}, {Teddy Seyed},
  {Xing-Dong Yang}, {and} {Bing-Yu Chen}. 2019.
\newblock \showarticletitle{Aarnio: {Passive} {Kinesthetic} {Force} {Output}
  for {Foreground} {Interactions} on an {Interactive} {Chair}}. In {\em
  Proceedings of the 2019 {CHI} {Conference} on {Human} {Factors} in
  {Computing} {Systems} - {CHI} '19}. ACM Press, Glasgow, Scotland Uk, 1--13.
\newblock
\showISBNx{978-1-4503-5970-2}
\showDOI{%
\url{http://dx.doi.org/10.1145/3290605.3300902}}


\bibitem{teng_pupop:_2018}
{Shan-Yuan Teng}, {Tzu-Sheng Kuo}, {Chi Wang}, {Chi-huan Chiang}, {Da-Yuan
  Huang}, {Liwei Chan}, {and} {Bing-Yu Chen}. 2018.
\newblock \showarticletitle{{PuPoP}: {Pop}-up {Prop} on {Palm} for {Virtual}
  {Reality}}. In {\em The 31st {Annual} {ACM} {Symposium} on {User} {Interface}
  {Software} and {Technology} - {UIST} '18}. ACM Press, Berlin, Germany, 5--17.
\newblock
\showISBNx{978-1-4503-5948-1}
\showDOI{%
\url{http://dx.doi.org/10.1145/3242587.3242628}}


\bibitem{teng_tilepop:_2019}
{Shan-Yuan Teng}, {Cheng-Lung Lin}, {Chi-huan Chiang}, {Tzu-Sheng Kuo}, {Liwei
  Chan}, {Da-Yuan Huang}, {and} {Bing-Yu Chen}. 2019.
\newblock \showarticletitle{{TilePoP}: {Tile}-type {Pop}-up {Prop} for
  {Virtual} {Reality}}.
\newblock  (2019), 11.
\newblock


\bibitem{teyssier_uduino_2019}
{Marc Teyssier}. 2019.
\newblock Uduino {\textbar} {Home}.
\newblock   (2019).
\newblock
\showURL{%
\url{https://marcteyssier.com/uduino/}}


\bibitem{usoh_walking_1999}
{Martin Usoh}, {Kevin Arthur}, {Mary~C. Whitton}, {Rui Bastos}, {Anthony
  Steed}, {Mel Slater}, {and} {Frederick~P. Brooks}. 1999.
\newblock \showarticletitle{Walking {\textgreater} walking-in-place
  {\textgreater} flying, in virtual environments}. In {\em Proceedings of the
  26th annual conference on {Computer} graphics and interactive techniques -
  {SIGGRAPH} '99}. ACM Press, Not Known, 359--364.
\newblock
\showISBNx{978-0-201-48560-8}
\showDOI{%
\url{http://dx.doi.org/10.1145/311535.311589}}


\bibitem{vonach_vrrobot:_2017}
{Emanuel Vonach}, {Clemens Gatterer}, {and} {Hannes Kaufmann}. 2017.
\newblock \showarticletitle{{VRRobot}: {Robot} actuated props in an infinite
  virtual environment}. In {\em 2017 {IEEE} {Virtual} {Reality} ({VR})}. IEEE,
  Los Angeles, CA, USA, 74--83.
\newblock
\showISBNx{978-1-5090-6647-6}
\showDOI{%
\url{http://dx.doi.org/10.1109/VR.2017.7892233}}


\bibitem{wang_movevr_2020}
{Yuntao Wang}, {Hanchuan Li}, {Zhengyi Cao}, {Huiyi Luo}, {Ke Ou}, {John
  Raiti}, {Chun Yu}, {Shwetak Patel}, {and} {Yuanchun Shi}. 2020.
\newblock \showarticletitle{{MoveVR}: {Enabling} {Multiform} {Force} {Feedback}
  in {Virtual} {Reality} using {Household} {Cleaning} {Robot}}.
\newblock  (2020), 12.
\newblock


\bibitem{wexelblat_virtual_1993}
{Alan Wexelblat}. 1993.
\newblock Virtual reality: applications and explorations.
\newblock   (1993).
\newblock
\showURL{%
\url{http://libertar.io/lab/wp-content/uploads/2016/02/Virtual.Reality.-.Applications.And_.Explorations.pdf/page=164}}
\newblock
\shownote{Myron Krueger, Artificial reality 2 An easy entry to Virtual reality
  Chap 7.}


\bibitem{whitmire_haptic_2018}
{Eric Whitmire}, {Hrvoje Benko}, {Christian Holz}, {Eyal Ofek}, {and} {Mike
  Sinclair}. 2018.
\newblock \showarticletitle{Haptic {Revolver}: {Touch}, {Shear}, {Texture}, and
  {Shape} {Rendering} on a {Reconfigurable} {Virtual} {Reality} {Controller}}.
  In {\em Proceedings of the 2018 {CHI} {Conference} on {Human} {Factors} in
  {Computing} {Systems} - {CHI} '18}. ACM Press, Montreal QC, Canada, 1--12.
\newblock
\showISBNx{978-1-4503-5620-6}
\showDOI{%
\url{http://dx.doi.org/10.1145/3173574.3173660}}


\bibitem{yafune_haptically_2011}
{M. Yafune} {and} {Y. Yokokohji}. 2011.
\newblock \showarticletitle{Haptically rendering different switches arranged on
  a virtual control panel by using an encountered-type haptic device}. In {\em
  2011 {IEEE} {World} {Haptics} {Conference}}. 551--556.
\newblock
\showDOI{%
\url{http://dx.doi.org/10.1109/WHC.2011.5945545}}


\bibitem{yamaguchi_non-grounded_2016}
{Kotaro Yamaguchi}, {Ginga Kato}, {Yoshihiro Kuroda}, {Kiyoshi Kiyokawa}, {and}
  {Haruo Takemura}. 2016.
\newblock \showarticletitle{A {Non}-grounded and {Encountered}-type {Haptic}
  {Display} {Using} a {Drone}}. In {\em Proceedings of the 2016 {Symposium} on
  {Spatial} {User} {Interaction} - {SUI} '16}. ACM Press, Tokyo, Japan, 43--46.
\newblock
\showISBNx{978-1-4503-4068-7}
\showDOI{%
\url{http://dx.doi.org/10.1145/2983310.2985746}}


\bibitem{ye_pull-ups:_2019}
{Yuan-Syun Ye}, {Hsin-Yu Chen}, {and} {Liwei Chan}. 2019.
\newblock \showarticletitle{Pull-{Ups}: {Enhancing} {Suspension} {Activities}
  in {Virtual} {Reality} with {Body}-{Scale} {Kinesthetic} {Force} {Feedback}}.
  In {\em Proceedings of the 32nd {Annual} {ACM} {Symposium} on {User}
  {Interface} {Software} and {Technology} - {UIST} '19}. ACM Press, New
  Orleans, LA, USA, 791--801.
\newblock
\showISBNx{978-1-4503-6816-2}
\showDOI{%
\url{http://dx.doi.org/10.1145/3332165.3347874}}


\bibitem{yokokohji_wysiwyf_1999}
{Yasuyoshi Yokokohji}, {Ralph~L. Hollis}, {and} {Takeo Kanade}. 1999.
\newblock \showarticletitle{{WYSIWYF} {Display}: {A} {Visual}/{Haptic}
  {Interface} to {Virtual} {Environment}}.
\newblock {\em Presence: Teleoperators and Virtual Environments\/} {8}, 4 (Aug.
  1999), 412--434.
\newblock
\showISSN{1054-7460, 1531-3263}
\showDOI{%
\url{http://dx.doi.org/10.1162/105474699566314}}


\bibitem{yokokohji_path_2001}
{Y. Yokokohji}, {J. Kinoshita}, {and} {T. Yoshikawa}. 2001.
\newblock \showarticletitle{Path planning for encountered-type haptic devices
  that render multiple objects in {3D} space}. In {\em Proceedings {IEEE}
  {Virtual} {Reality} 2001}. 271--278.
\newblock
\showDOI{%
\url{http://dx.doi.org/10.1109/VR.2001.913796}}


\bibitem{zielasko_either_2020}
{Daniel Zielasko} {and} {Bernhard~E Riecke}. 2020.
\newblock \showarticletitle{Either {Give} {Me} a {Reason} to {Stand} or an
  {Opportunity} to {Sit} in {VR}}.
\newblock  (2020), 3.
\newblock


\end{thebibliography}

\end{document}